# Atom-surface physics: A review

Athanasios Laliotis[1], Bing-Sui Lu[2], Martial Ducloy[1*], and David Wilkowski[2,3,4,5]

[1]Laboratoire de Physique des Lasers, UMR 7538 du CNRS, Université Paris13-Sorbonne-Paris-Cité, F-93430 Villetaneuse, France.

[2]School of Physical and Mathematical Sciences, Nanyang Technological University, 21 Nanyang Link, Singapore 637371, Singapore.

[3]Centre for Disruptive Photonic Technologies, TPI & SPMS, Nanyang Technological University, 637371 Singapore, Singapore.

[4]MajuLab, International Joint Research Unit UMI 3654, CNRS, Université Cote d'Azur, Sorbonne Université, National University of Singapore, Nanyang Technological University, Singapore, Singapore.

[5]Centre for Quantum Technologies, National University of Singapore, 117543 Singapore, Singapore.

* Corresponding author: martial.ducloy@univ-paris13.fr

**Abstract:**

**An atom in front of a surface is one of the simplest and fundamental problem in physics. Yet, it allows testing quantum electrodynamics, while providing potential platforms and interfaces for quantum technologies. Despite, its simplicity, combined with strong scientific and technological interests, atom-surface physics, at its fundamental level, remains largely unexplored mainly because of challenges associated with precise control of the atom-surface distance. Nevertheless, substantial breakthroughs have been made over the last two decades. With the development of cold and quantum atomic gases, one has gained further control on atom-surface position, naturally leading to improved precision in the Casimir-Polder interaction measurement. Advances have also been reported in finding experimental *knobs* to tune and even reverse the Casimir-Polder interaction strength. So far, this has only been achieved for atoms in short-lived excited states, however, the rapid progresses in material sciences, e.g. metamaterials and topological materials have inspired new ideas for controlling the atom-surface interaction in long-lived states. In addition, combining nano-photonic and atom-surface physics is now envisioned for applications in quantum information processing. The first purpose of this review is to give a general overview on the latest experimental developments in atom-surface physics. The second main objective is to sketch a vision of the future of the field, mainly inspired by the abundant theoretical works and proposals available now in the literature.**

# Table of contents

## 1. Introduction

Among the few macroscopic manifestations of quantum mechanics (QM), like Bose-Einstein condensation, superconductivity or superfluidity, those linked to the zero-point energy in quantum mechanics and the associated quantum fluctuations of the electromagnetic (EM) field were predicted by Casimir in 1948 [1]. The presence of EM quantum fluctuations leads to an attractive Casimir force between two perfectly conducting plates facing each other in vacuum. The same year, Casimir together with Polder [2] analyzed the interaction of a neutral, polarizable, microscopic quantum mechanical system (*e.g.* a neutral ground state atom or molecule) with a reflecting plane surface – an archetype of a microscopic system interacting with a macroscopic body. Their work highlighted the importance of the EM fluctuations in vacuum accounting for retardation effects due to the finite speed of light. The quantum electrodynamics (QED) approach of Casimir and Polder shed new light on the problem of atom-surface interactions, going beyond the electrostatic description of a dipole interacting with its surface-induced image that was proposed by Lennard-Jones in 1932 [3]. This ground-breaking theoretical work [2] was later generalized to account for the dielectric response of the material, the influence of thermal fluctuations as well as short-lived excited state atoms, [4], [5] [6], [7], [8].

The first precision measurements of atom-surface interactions were achieved more than forty years after Casimir's theory, using thermal beams of alkali atoms [9], [10] or selective reflection spectroscopy in alkali vapor cells [11], [12]. Around the same time, the development of laser cooling and trapping also opened new perspectives in Casimir-Polder (CP) measurement [13]. Experiments based on cold atoms or ions have flourished over the last twenty years, allowing CP precision measurements from the nano [14], the micro [15], and up to macroscopic distances [16]. In addition, the influence of thermal fluctuations in CP interactions was also experimentally investigated [15], [17], following the remarkable realization of thermal field coherence at the vicinity of surfaces [18], [19] and booming studies of near-field energy transfer [20]. Finally, the 21$^{st}$ century has brought significant advances in the field of nanofabrication, nanophotonics and material sciences. These remarkable new technologies have allowed a new generation of hybrid systems putting atoms in close proximity to solid state platforms [21] thus making the possibility of controlling atom-surface interactions a key issue in CP physics.

A major part of this review focuses on the latest developments in CP precision measurement and in new exciting prospects for tuning atom-surface interactions using new materials. At first, in section 2, we present an overview of theoretical developments in CP theory. The theoretical works based on the quantum-mechanical linear-response formalism and Green function approach to both atom and EM field susceptibilities are discussed in detail. The influence of surface thermal excitations at non-zero temperature is also considered. Both CP-induced atomic energy level shifts and level decay rates are reviewed. We are also considering the short distance limit, namely the non-retarded regime, also called the van der Waals regime, where complete analytical derivations are possible.

In Section 3, we describe the various experimental approaches used to study the atom-surface interactions. Historically the first experiments were based on the mechanical deflection of thermal atomic beams by the CP forces exerted by plane or cylindrical materials. Later, those studies were extended with ultracold atom. Another approach was simultaneously developed, based on spectral measurements of the CP-induced atomic transition frequency shifts in confined atomic vapors, via either selective reflection spectroscopy or spectroscopic probing of thin vapor cells. In an alternative manner, interferometric measurements in material nanogratings were developed from the 2000's, giving access to the phase shift induced on atomic matter waves by the atom-surface potential.

In Section 4, various ways of using surface plasmon/phonon polariton modes to tune CP interactions and make them resonant with atomic transitions are explored. One approach relies on temperature tuning of the material surface in order to control the level of thermal excitations of the surface mode. An alternative way is based upon specially designed nanostructured metasurfaces to control the surface plasmon frequency and tune it around atomic resonances.

Section 5 focuses on new materials, which could be potentially used as interfaces: material nanostructures and metamaterials, magnetic materials, and topological materials. The importance of anisotropic materials in the search for CP repulsion is underlined.

Finally, in section 6, we give an outlook of the field. We envision the growing importance of nanotechnologies and topological material to engineer atom-surface interaction. Moreover, we point out the emergence of new type of quantum emitters such as molecules and Rydberg atoms to explore CP interaction beyond the electric dipole approximation.

## 2. Theory background

In this section, we summarize the theoretical methods to address electromagnetic (EM) mediated interaction of an atom with a solid object at its vicinity. More precisely, how the energy levels of the atom are shifted, and how the excited state lifetimes are modified. A breakthrough contribution came, in 1948, from Casimir and Polder who derived the atom-wall interaction, accounting for the retardation effects [2]. A few years later, the theory was reconsidered using modern quantum electrodynamics (QED) [5]; subsequently, the linear-response formalism was applied to an arbitrary flat interface; see for example [7]. More recently, this approach has been used to explore dielectric [22] or metallic slabs, new materials with exotic permittivity, like hyperbolic materials, or two-dimensional materials, such as graphene or Chern insulators (see section 5 for more details). Other geometrical configurations such as spheres or cylinders were also considered. In the large size limit, the CP interaction weakly depends on the object geometry [23]–[25]. However, with the recent advances of nanophotonics, it is now possible to precisely design objects with sizes comparable to the characteristic resonant wavelength of the atom [26]. Here, CP interaction is strongly modified [27]. These nano-objects can be isolated or arranged in periodic manner, to form bulk or surface metamaterials and can be used to tune the atom-surface interaction (see Sections 4 and 5). Some of these nanophotonics devices behave as open optical resonators, characterized by a limited number of quasi-normal modes. Here, the main motivation is to achieve strong coupling between a quantum emitter and one of the quasi-normal mode [28]. Theoretical predictions are performed using QED and linear-response formalism, with a challenging task to properly compute the spatial extension of the quasi-normal mode [29]. A detailed discussion of this issue, which is still an active subject of research, is beyond the scope of the present review, and we limit ourselves to highlight the common theoretical approach for both CP interaction and nanophotonics.

### 2.1 Casimir-Polder energy shift

We now calculate how the energy levels and emission rate of the atom are modified by the presence of surrounding materials. The atom is here considered as a point source (electric dipole approximation), characterized by optical frequencies $\omega_{mk} = (E_m - E_k)/\hbar$ with electric dipole moments $\boldsymbol{d}_{mk}$. First, let us consider the CP shift of the energy level of an atomic state. This arises due to the interaction of the atom with radiation, and can be calculated using second-order perturbation theory. We assume that the system consists of an atom in a given state interacting with radiation, which exists in an ensemble of thermal states. In the absence of atom-radiation interaction, a given eigenstate of the system can be expressed as $|i\rangle \otimes |I\rangle$ (where $|i\rangle$ and $|I\rangle$ are respectively the atomic state and the radiation field state), which obeys

$$H_0|i\rangle \otimes |I\rangle = (E_i + E_I)|i\rangle \otimes |I\rangle. \qquad 2.1$$

Here $E_i$ and $E_I$ are the energies of the atomic state $|i\rangle$ and radiation field state $|I\rangle$, respectively. In the presence of a weakly perturbing interaction $H_I$, the system's overall eigenstate is modified to

$$|\psi\rangle = |i\rangle \otimes |I\rangle + \sum_{k,K}{}' (\lambda c_{kK}^{(1)} + \lambda^2 c_{kK}^{(2)})|k\rangle \otimes |K\rangle, \qquad 2.2$$

where the primed symbol means that the states $|k\rangle \otimes |K\rangle$ are distinct from $|i\rangle \otimes |I\rangle$. The energy equation becomes

$$(H_0 + \lambda H_I)|\psi\rangle = \left(E_i + E_I + \lambda \delta E^{(1)} + \lambda^2 \delta E^{(2)}\right)|\psi\rangle, \qquad 2.3$$

where $\lambda^n$ denotes the *n*th order of the perturbation expansion, $c_{kK}^{(1)}$ and $c_{kK}^{(2)}$ are respectively the first- and second-order perturbation corrections to the eigenstate of the system, and $\delta E^{(1)}$ and $\delta E^{(2)}$ are the corresponding corrections to the eigenenergy of the system. At first order in the perturbation expansion, we find that $\delta E^{(1)} = 0$ and

$$c_{iI}^{(1)} = \frac{\langle I| \otimes \langle i|H_I|k\rangle \otimes |K\rangle}{E_k + E_K - E_i - E_I}. \qquad 2.4$$

At second order in the perturbation expansion, we find

$$\delta E^{(2)} = \sideset{}{'}\sum_{k,K} c_{kK}^{(1)} \langle I| \otimes \langle i|H_I|k\rangle \otimes |K\rangle = \sideset{}{'}\sum_{k,K} \frac{|\langle I| \otimes \langle i|H_I|k\rangle \otimes |K\rangle|^2}{E_i + E_I - E_k - E_K}. \qquad 2.5$$

This energy shift is obtained assuming that the unperturbed radiation field state is $|I\rangle$, which occurs with a probability of $P(I)$. As there is an ensemble of such states, we need to average the energy shift over the entire ensemble, after which we obtain the shift in the (free) energy of the atomic state $|i\rangle$:

$$\Delta E_i = \sum_I P(I) \delta E^{(2)} = \frac{1}{\hbar} \mathbf{P} \sum_{I,K,k} P(I) \frac{|\langle I| \otimes \langle i|H_I|k\rangle \otimes |K\rangle|^2}{\omega_i + \omega_I - \omega_k - \omega_K}, \qquad 2.6$$

where $\mathbf{P}$ denotes taking the principal value. For an electric dipole in vacuum $H_I = -\mathbf{d} \cdot \boldsymbol{\mathcal{E}}$, where $\mathbf{d}$ is the dipole operator and $\boldsymbol{\mathcal{E}}$ is the electric field operator. The free energy shift of the atomic state $|i\rangle$ becomes [32]

$$\Delta E_i = \frac{1}{\hbar} \mathbf{P} \sum_{I,K,k} P(I) \frac{\mathcal{E}_{IK,\alpha}(\mathbf{r})\mathcal{E}_{KI,\beta}(\mathbf{r}) d_{ik,\alpha} d_{ki,\beta}}{\omega_i + \omega_I - \omega_k - \omega_K}, \qquad 2.7$$

where $\mathcal{E}_{IK,\alpha}(\mathbf{r}) \equiv \langle I|\mathcal{E}_\alpha(\mathbf{r})|K\rangle$ is the transition matrix element for the electric field operator, $d_{ik,\alpha} \equiv \langle i|d_\alpha|k\rangle$ is the electric dipole transition matrix element, $\alpha = x, y, z$ are the Cartesian indices, and $\mathbf{r}$ is the position vector of the atom. In the above, we have used the electric field operator in the Schrödinger picture, which is related to the interaction-picture field operator $\mathcal{E}_\alpha(\mathbf{r}, t)$ via $\mathcal{E}_\alpha(\mathbf{r}, t) = e^{iH_I t/\hbar} \mathcal{E}_\alpha(\mathbf{r}) e^{-iH_I t/\hbar}$. The electric field is related to the so-called dyadic Green function $G_{\alpha\beta}(\mathbf{r}, \mathbf{r}'; t)$ via

$$G_{\alpha\beta}(\mathbf{r}, \mathbf{r}'; t) = \frac{i}{\hbar} \langle [\mathcal{E}_\alpha(\mathbf{r}, t), \mathcal{E}_\beta(\mathbf{r}', 0)] \rangle \theta(t), \qquad 2.8$$

where the angular brackets denote ensemble averaging at the given temperature, and $\theta(t)$ is a Heaviside step function with the value of 1 (0) if $t > 0$ ($t < 0$). The dyadic Green function $\mathbf{G}(\mathbf{r}, \mathbf{r}', \omega)$, essentially gives the electric field at $\mathbf{r}$ emitted by a point source at a position $\mathbf{r}'$ [33]. Importantly, the fluctuation-dissipation theorem in the linear-response theory enables one to relate the imaginary part of the dyadic Green function, which is a classical quantity, to the electric field correlation function $\mathbf{C}(\mathbf{r}, \mathbf{r}', \omega) = \int dt \langle \boldsymbol{\mathcal{E}}(\mathbf{r}, t) \boldsymbol{\mathcal{E}}(\mathbf{r}', 0) \rangle e^{i\omega t}$ [7], [8], [32], which can be computed in the framework of quantum electrodynamics, in which vacuum fluctuations are properly taken into consideration. When material surfaces are present, the dyadic Green function can be decomposed in the following manner,

$$\mathbf{G}(\mathbf{r}, \mathbf{r}', \omega) = \mathbf{G}^{(0)}(\mathbf{r}, \mathbf{r}', \omega) + \mathbf{G}^{(sc)}(\mathbf{r}, \mathbf{r}', \omega), \qquad 2.9$$

where $\mathbf{G}^{(0)}(\mathbf{r}, \mathbf{r}', \omega)$ is the dyadic Green function in free space and $\mathbf{G}^{(sc)}(\mathbf{r}, \mathbf{r}', \omega)$ is the scattering term taking into consideration the spatial structure of the dielectric permittivity and magnetic permeability. By making use of linear-response theory, the details of which are presented fully in Refs. [16] and [17], one finds that there are resonant and non-resonant contributions to the energy shift of the state $|m\rangle$:

$$\Delta E_m(\mathbf{r}) = \Delta E_m^{(r)}(\mathbf{r}) + \Delta E_m^{(or)}(\mathbf{r}), \qquad 2.10$$

where $\Delta E_m^{(r)}$ denotes the resonant contribution and $\Delta E_m^{(or)}$ denotes the off-resonant contribution. The temperature dependence of the non-resonant term of the CP interaction has been studied since

Lifshitz [4] and McLachlan [6], while complete treatments of the thermal effects, including resonant interactions were presented much later [32], [35], [36].

### 2.1.1 Nonzero temperature

For nonzero temperature, besides downward atomic transitions $|m\rangle \to |k\rangle$ (where $\omega_{km} < 0$), there can also be upward atomic transitions $|m\rangle \to |k\rangle$ (where $\omega_{km} > 0$) due to excitation by thermal photons. Taking these two types of transitions into account, the resonant contribution is given by [32]:

$$\Delta E_m^{(r)}(\mathbf{r}) = \mu_0 \sum_{k>m} \omega_{km}^2 \mathrm{n}(\omega_{km}, T) \mathbf{d}_{mk} \mathrm{Re}\{\mathbf{G}^{(sc)}(\mathbf{r},\mathbf{r},\omega_{km})\}\mathbf{d}_{km} \qquad 2.11$$
$$-\mu_0 \sum_{k<m} \omega_{km}^2 [1 + \mathrm{n}(\omega_{mk}, T)] \mathbf{d}_{mk} \mathrm{Re}\{\mathbf{G}^{(sc)}(\mathbf{r},\mathbf{r},\omega_{mk})\}\mathbf{d}_{km}.$$

Here, $\mu_0 = 4\pi \times 10^{-7}$ H/m is the vacuum permeability, $\mathbf{d}_{km}$ is the matrix element for a dipole transition from $|m\rangle \to |k\rangle$, and $\mathrm{n}(\omega, T) = \left(e^{\hbar\omega/k_B T} - 1\right)^{-1}$ is the mean occupation number according to Bose-Einstein statistics at the transition frequency.

The off-resonant term is expressed in terms of a summation over the Matsubara frequencies, $\xi_p = 2\pi \frac{k_B T}{\hbar} p$ that is given by [32], [26]

$$\Delta E_m^{(or)}(\mathbf{r}) = \frac{2\mu_0 k_B T}{\hbar} \sum_k \sum_{p=0}^{\infty}{}' \frac{\omega_{mk} \xi_p^2}{\omega_{mk}^2 + \xi_p^2} \mathbf{d}_{mk} \mathbf{G}^{(sc)}(\mathbf{r},\mathbf{r},i\xi_p) \mathbf{d}_{km}. \qquad 2.12$$

Here, the prime symbol signifies that the first term of the sum should be multiplied by 1/2. The symbol $\mathbf{G}^{(sc)}(\mathbf{r},\mathbf{r},\omega)$ denotes the scattering Green tensor for a dipole of frequency $\omega$ interacting with its own reflected field ($\mathbf{r}' = \mathbf{r}$). For a single dielectric interface that preserves Lorentz reciprocity, the scattering Green tensor is given by [38]

$$\mathbf{G}^{(sc)}(\mathbf{r},\mathbf{r},\omega) = \frac{i}{8\pi} \int_0^{\infty} dq \, \frac{q \, e^{2i\sqrt{(\omega/c)^2 - q^2}\,z}}{\sqrt{(\omega/c)^2 - q^2}} \left( r_s \begin{pmatrix} 1 & 0 & 0 \\ 0 & 1 & 0 \\ 0 & 0 & 0 \end{pmatrix} \right. \qquad 2.13$$
$$\left. + \frac{r_p c^2}{\omega^2} \begin{pmatrix} q^2 - (\omega/c)^2 & 0 & 0 \\ 0 & q^2 - (\omega/c)^2 & 0 \\ 0 & 0 & 2q^2 \end{pmatrix} \right),$$

where

$$r_s = \frac{\mu\sqrt{\left(\frac{\omega}{c}\right)^2 - q^2} - \sqrt{\varepsilon\mu\left(\frac{\omega}{c}\right)^2 - q^2}}{\mu\sqrt{\left(\frac{\omega}{c}\right)^2 - q^2} + \sqrt{\varepsilon\mu\left(\frac{\omega}{c}\right)^2 - q^2}}, \qquad 2.14a$$

$$r_p = \frac{\varepsilon\sqrt{\left(\frac{\omega}{c}\right)^2 - q^2} - \sqrt{\varepsilon\mu\left(\frac{\omega}{c}\right)^2 - q^2}}{\varepsilon\sqrt{\left(\frac{\omega}{c}\right)^2 - q^2} + \sqrt{\varepsilon\mu\left(\frac{\omega}{c}\right)^2 - q^2}} \qquad 2.14b$$

are respectively the reflection coefficients for the s- and p-polarized waves. Equations (2.11) and (2.12) give the general form of the energy shift of an atomic state. In the remaining part of this section, we will give practical examples in various regimes.

#### 2.1.1.1 Near-field: The non-retarded regime

Now we turn to the near-field limit of the atom-surface interaction, which corresponds to $z \ll \frac{\lambda_{mk}}{4\pi}$ (or equivalently, $z\sqrt{|\varepsilon\mu|}|\omega_{mk}|/c \ll 1$), where $z$ is the distance of the atom above the surface and $\lambda_{mk}$ is the wavelength of the $|m\rangle \to |k\rangle$ dipole coupling. Let us consider a configuration in which the half-space

$z < 0$ is filled up with a non-magnetic and homogeneous material. In the near-field limit, the scattering dyadic Green function in Eq. (2.13) takes the following simple form [38]

$$\boldsymbol{G}^{(sc)}(\boldsymbol{r},\boldsymbol{r},\omega) = \frac{c^2}{32\pi\omega^2 z^3}\frac{\varepsilon(\omega)-1}{\varepsilon(\omega)+1}\begin{pmatrix} 1 & 0 & 0 \\ 0 & 1 & 0 \\ 0 & 0 & 2 \end{pmatrix}. \qquad 2.15$$

It is convenient to decompose the resonant energy shift into an upward channel contribution and a downward channel contribution. Using Eqs. (2.11), (2.12) and (2.15), one finds

$$\Delta E_m(\boldsymbol{r}) = \frac{2h}{z^3}\sum_{k>m} C_3^{m\to k} n(\omega_{km},T) Re\left[\frac{\varepsilon(\omega_{km})-1}{\varepsilon(\omega_{km})+1}\right] \qquad 2.16$$
$$- \frac{2h}{z^3}\sum_{k<m} C_3^{m\to k}[n(\omega_{mk},T)+1] Re\left[\frac{\varepsilon(\omega_{mk})-1}{\varepsilon(\omega_{mk})+1}\right]$$
$$- \frac{8\pi k_B T}{z^3}\sum_k \sum_{p=0}^{\infty}{}' C_3^{m\to k} \frac{\omega_{mk}}{\omega_{mk}^2+\xi_p^2}\frac{\varepsilon(i\xi_p)-1}{\varepsilon(i\xi_p)+1},$$

where the first term represents the upward channel contribution, the second term represents the downward channel contribution, and the third term represents the off-resonant contribution. Here, we have introduced the so-called $C_3$ coefficient, defined for transition $|m\rangle \to |k\rangle$ by [12]

$$C_3^{m\to k} \equiv \frac{1}{64\pi\varepsilon_0 h}[|\boldsymbol{d}_{mk}|^2 + (\boldsymbol{d}_{mk}\cdot\hat{z})^2]. \qquad 2.17$$

We thus see that in the near field, both terms follow the van der Waals law, having a distance dependence of $z^{-3}$, and the $C_3$ coefficient quantifies the strength of the van der Waals interaction with a perfect conductor. The situation corresponds to the non-retarded CP interaction (also called van der Waals interaction) where the atom-interface distance $z$ is sufficiently small such that the phase rotation of the field during propagation can be disregarded. It is common to consider an isotropic atom (an average over all m quantum numbers) [39]. In this case we find that the van der Waals coefficient of the $|m\rangle \to |k\rangle$ coupling for a perfect conductor is given by:

$$C_3^{m\to k} \sim A_{mk}|\omega_{mk}|^{-3}, \qquad 2.18$$

where $A_{mk}$ is the transition probability of the $|m\rangle \to |k\rangle$ coupling. Transition probability tables for alkali atoms can be found in [40], [41], [42].

### 2.1.1.2 Far-field: Retardation effect

The resonant and non-resonant terms of the CP interaction have significantly different behaviors as a function of distance. The non-resonant term distance dependence decays rather quickly from a $z^{-3}$ to a $z^{-4}$ regime, first studied by Casimir and Polder in [2]. Retardation effects of the non-resonant CP contribution have been extensively studied theoretically [43], [32], and experimentally with ground state alkali atoms [10], [13], [14] for distances ranging from $\sim 50-500$ nm. In the so-called Lifshitz regime [4], where the separation distance $z$ is greater than the thermal wavelength $\lambda_T = \hbar c/k_B T$ (typically $\sim 7$ µm at room temperature) for regular dielectric materials, the non-resonant contribution falls back to a $z^{-3}$ distance dependence that has been predicted theoretically [6], [44] but never demonstrated experimentally [15]. In the Lifshitz regime, only the zero Matsubara frequency term contributes to the scattering Green tensor as the contributions with nonzero Matsubara frequencies are exponentially suppressed, and thus the non-resonant CP interaction becomes

$$\Delta E_m^{(or)}(\boldsymbol{r}) = -\frac{4\pi}{z^3}\sum_k C_3^{m\to k}\frac{k_B T}{\omega_{mk}}\frac{\varepsilon(0)-1}{\varepsilon(0)+1}. \qquad 2.19$$

The resonant term of the CP interaction decays more slowly with distance [7], [8], [43], [45] compared to the off-resonant term discussed above. In the far-field regime $z \gg \frac{\lambda_{mk}}{4\pi}$, the resonant term oscillates from repulsion to attraction with a periodicity of $\lambda_{mk}/2$, a typical behavior of a classical oscillating dipole that interacts with its own reflected field [43]. In the far-field approximation, the distance dependence of the $|m\rangle \to |k\rangle$ dipole contribution, $\Delta E_{mk}^{(r)}(\boldsymbol{r})$, on the resonant CP term becomes [45]:

$$\Delta E_{mk}^{(r)}(\boldsymbol{r}) \propto -\frac{1}{z}|\rho(\omega_{mk})|\cos\left(\frac{4\pi}{\lambda_{mk}}z + \phi(\omega_{mk})\right), \qquad 2.20$$

where $\rho(\omega_{mk})$ is the reflection coefficient of the surface at the transition frequency and $\phi(\omega_{mk})$ is its argument. The temperature dependence of the resonant term is given in Eq. (2.11) depending on the nature of the coupling (absorption or emission).

### 2.1.2 Zero temperature

The preceding formulas are valid for atomic transition frequencies for which $\hbar\omega$ is comparable to or less than $k_B T$. For optical frequencies, $\hbar\omega/k_B T \gg 1$, and the effect of temperature can be ignored. At zero temperature, the upward couplings are absent, and the resonant CP energy shift becomes [32], [26], [31], [46]

$$\Delta E_m^{(r)}(\boldsymbol{r}) = -\mu_0 \sum_{k<m} \omega_{mk}^2 \boldsymbol{d}_{mk} \text{Re}\{\boldsymbol{G}^{(sc)}(\boldsymbol{r},\boldsymbol{r},\omega_{mk})\}\boldsymbol{d}_{km}. \qquad 2.21$$

The off-resonant contribution at zero temperature reads [32], [26], [31], [46]

$$\Delta E_m^{(or)}(\boldsymbol{r}) = -\frac{\mu_0}{\pi}\sum_k \omega_{mk}\int_0^\infty d\xi \frac{\xi^2}{\omega_{mk}^2+\xi^2}\boldsymbol{d}_{mk}\boldsymbol{G}^{(sc)}(\boldsymbol{r},\boldsymbol{r},i\xi)\boldsymbol{d}_{km}. \qquad 2.22$$

Eqs. (2.21) and (2.22) are obtained using second-order perturbation theory and making use of the linear response theory and the fluctuation-dissipation theorem, as described earlier [47], [7], [8], [38], [48]. The latter integral in Eq. (2.22) is performed along the positive imaginary frequency axis, which does not have poles. Singularities due to atomic resonance are taken into account in the resonant term corresponding to Eq. (2.21). The resonant part comes from a resonant energy transfer between the atom and the material surface, which occurs when the atom has a de-excitation frequency matching one of the absorption frequencies of the surface, and a real photon emitted by the atom then gets absorbed by the surface. In the ground state of the atom, only the off-resonant term is non-zero. It comes from the fluctuation of the dipole moment induced by the vacuum and has a clear quantum origin [38]. The dyadic Green function in free space gives also a non-zero contribution which is the celebrated Lamb shift [49]. Since the vacuum Lamb shift gives a position-independent energy shift, it is included in the definition of the atomic energies. We note that the dyadic Green function in Eqs (2.21) and (2.22) needs to be evaluated at the atom position where both emission and field positions are merged.

#### 2.1.2.1 Near-field: Non-retardation limit

Using Eqs. (2.15), (2.21) and (2.22), the energy shift reads,

$$\Delta E_m^{(r)}(\boldsymbol{r}) = -\frac{1}{32\pi\varepsilon_0\omega^2 z^3}\sum_{k<m}\omega_{mk}^2 \text{Re}\left\{\frac{\varepsilon(\omega_{mk})-1}{\varepsilon(\omega_{mk})+1}\right\}[|\boldsymbol{d}_{mk}|^2 + (\boldsymbol{d}_{mk}\cdot\hat{z})^2], \qquad 2.23a$$

$$\Delta E_m^{(or)}(\boldsymbol{r}) = -\frac{1}{32\pi^2\varepsilon_0 z^3}\sum_k \omega_{mk}[|\boldsymbol{d}_{mk}|^2 + (\boldsymbol{d}_{mk}\cdot\hat{z})^2]\int_0^{+\infty}d\xi \frac{1}{\omega_{mk}^2+\xi^2}\frac{\varepsilon(i\xi)-1}{\varepsilon(i\xi)+1}, \qquad 2.23b$$

for the resonant and off-resonant contribution, respectively.

Those expressions trigger several important comments. First, we note the energies follow a $z^{-3}$ position dependence, as it is expected for a dipole interacting with its own image though the material [50]. Second, the off-resonant term is negative leading to the usual attractive Van der Waals force for an atom in the ground state. However, one may have $|\varepsilon(\omega_{mk})| < 1$, indicating a repulsive resonant CP interaction that might become the dominant term in the excited state. This effect occurs if surface resonant modes in the material have their frequency at the vicinity of an atomic resonance (see section 4). Finally, we note that the dyadic Green function is not isotropic leading to mechanical action that depends on the dipole orientation [51], and atomic level mixing [52].

## 2.2 Decay rate

Similar to the CP energy shift, the decay rate $\Gamma_m(\boldsymbol{r})$ of an atomic state $|m\rangle$ can also be expressed in terms of the dyadic Green function. The decay rate is expressed by [38]

$$\Gamma_m(\boldsymbol{r}) = \sum_k \Gamma_{mk}(\boldsymbol{r}), \qquad 2.24$$

where for the general case of nonzero temperature, all allowed dipole couplings are included in the summation, including upward couplings which are due to the presence of thermal photons. (At zero temperature, the upward couplings would be absent). The decay rate for an upward channel $|m\rangle \to |k\rangle$ (where $\omega_{mk} < 0$) reads [53] [35], [32], [36]

$$\Gamma_{mk}(\boldsymbol{r}) = \frac{2\mu_0}{\hbar} \omega_{mk}^2 \boldsymbol{d}_{mk} \mathrm{n}(\omega_{km}, T) \mathrm{Im}\{\boldsymbol{G}(\boldsymbol{r}, \boldsymbol{r}, \omega_{km})\} \boldsymbol{d}_{km} \Theta(\omega_{km}). \qquad 2.25$$

In the case of a downward coupling (where $\omega_{mk} > 0$) the decay rate now reads

$$\Gamma_{mk}(\boldsymbol{r}) = \frac{2\mu_0}{\hbar} \omega_{mk}^2 \boldsymbol{d}_{mk}[1 + \mathrm{n}(\omega_{mk}, T)] \mathrm{Im}\{\boldsymbol{G}(\boldsymbol{r}, \boldsymbol{r}, \omega_{mk})\} \boldsymbol{d}_{km} \Theta(\omega_{mk}). \qquad 2.26$$

### 2.2.1 Zero temperature

As in the case of the CP energy shift, the above formulas can be approximated by their zero temperature limits for optical frequencies. For such a case, upward transition channels are absent, and the decay rate in Eq. (2.26) becomes

$$\Gamma_m(\boldsymbol{r}) = \sum_{k<m} \Gamma_{mk}(\boldsymbol{r}), \qquad 2.27$$

where

$$\Gamma_{mk}(\boldsymbol{r}) = \frac{2\mu_0}{\hbar} \omega_{mk}^2 \boldsymbol{d}_{mk} \mathrm{Im}\{\boldsymbol{G}(\boldsymbol{r}, \boldsymbol{r}, \omega_{mk})\} \boldsymbol{d}_{km}. \qquad 2.28$$

Since we consider $T = 0$, the vacuum modes are not populated, and the summation in Eq. (2.26) is done over the energy levels located below the level $m$ in consideration. Moreover, the dyadic Green function depends only on the atomic frequency $\omega_{mk}$, where the free space component gives the usual natural decay rate in vacuum, and the scattering term its modification due to the presence of the material. More precisely, the local density of state $\rho(\boldsymbol{r}, \omega_{mk})$ that enters in the derivation of the state decay rate $\Gamma_{mk} = \pi d_{mk}^2 \omega_{mk} \rho(\boldsymbol{r}, \omega_{mk})/(3\hbar\varepsilon_0)$, using the Fermi-golden rule, reads [33], [54], [55]: $\rho(\boldsymbol{r}, \omega_{mk}) = \frac{6\omega_{mk}}{\pi c^2} \boldsymbol{u}_{mk}^* \mathrm{Im}\{\boldsymbol{G}(\boldsymbol{r}, \boldsymbol{r}, \omega_{mk})\} \boldsymbol{u}_{mk}$. $\boldsymbol{u}_{mk} = \boldsymbol{d}_{mk}/|\boldsymbol{d}_{mk}|$ is the unitary vector along the direction of the dipole moment.

#### 2.2.1.1 Near-field: Non-retardation limit

In the near-field limit, the fluorescence decay rate of the atomic dipole can be approximated by

$$\Gamma_m^{(r)}(\boldsymbol{r}) = \frac{1}{16\pi\varepsilon_0 \hbar z^3} \sum_{k<m} \mathrm{Im}\left\{\frac{\varepsilon(\omega_{mk}) - 1}{\varepsilon(\omega_{mk}) + 1}\right\} [|\boldsymbol{d}_{mk}|^2 + (\boldsymbol{d}_{mk} \cdot \hat{z})^2]. \qquad 2.29$$

This equation describes the modification of the decay rate due to the presence of evanescent surface mode, and was experimentally demonstrated in cesium vapor cell [56].

## 2.3 Open optical resonators

In the previous subsections, we have seen how the modification of the dipolar emission and the CP energy shift of an atom at the vicinity of a material can be evaluated using the dyadic Green function. In general, analytical expressions of the dyadic Green function are not available [33], [55], and the development of numerical methods to address this problem is still an active research subject [57].

However, in some cases relevant here, the dyadic function can be expressed in a simple form, like for the CP interaction in the non-retarded limit. In the presence of quasi-normal modes, no general expression exists, but one can use a spectral representation that gives interesting insights and allows one to compute relevant quantities, such as the Purcell factor.

As an example where the quasi-normal modes approach is used, we consider open optical resonators (which can be nanorods or nanospheres). Here the quasi-normal modes approach is used since the system is characterized by resonances as well as field relaxation [58], [59]. If a few quasi-normal modes are enough to describe the response of the system, it is convenient to use the spectral representation of the dyadic Green function [60], [61],

$$\boldsymbol{G}^{(sc)}(\boldsymbol{r},\boldsymbol{r}',\omega) = c^2 \sum_\alpha \frac{\mathcal{E}_\alpha(\boldsymbol{r}) \otimes \mathcal{E}_\alpha(\boldsymbol{r}')}{2\omega(\omega_\alpha - \omega)}. \qquad 2.30$$

$\mathcal{E}_\alpha(\boldsymbol{r})$ are the fields of the quasi-normal modes. The sum is over all the relevant modes of complex frequency $\omega_\alpha = \omega'_\alpha + i\omega''_\alpha$. The imaginary part, $\omega''_\alpha < 0$, takes in account the radiative and non-radiative losses. The quality factor of the resonance is given by $Q = -\omega'_\alpha/2\omega''_\alpha$, and it is moderate for such system ($Q \sim 10$). Using Eqs. (2.21) and (2.28), the decay rate and the energy shift for the case of zero temperature read, respectively,

$$\Gamma_m^{(r)}(\boldsymbol{r}) = \frac{1}{\varepsilon_0 \hbar} \sum_{k<m} \left[ \omega_{mk}^2 |\boldsymbol{d}_{km}|^2 \sum_\alpha \text{Im}\left\{ \frac{1}{V_\alpha(\omega - \omega_\alpha)} \right\} \right], \qquad 2.31a$$

$$\Delta E_m^{(r)}(\boldsymbol{r}) = -\frac{1}{2\varepsilon_0} \sum_{k<m} \left[ \omega_{mk}^2 |\boldsymbol{d}_{km}|^2 \sum_\alpha \text{Re}\left\{ \frac{1}{V_\alpha(\omega - \omega_\alpha)} \right\} \right], \qquad 2.31b$$

where $V_\alpha = |\boldsymbol{d}_{km}|^2/[\boldsymbol{d}_{km} \cdot \boldsymbol{E}_\alpha(\boldsymbol{r})]^2$ is the mode volume [58].

# 3. Casimir-Polder measurement

The measurement of atom-surface interactions can be divided in three main approaches:

(1) The mechanical approach that monitors the influence of CP potential and forces on the external motion of the atom center-of-mass.

(2) The diffraction and interferometry approach that monitors diffraction patterns of atomic and molecular wave-packets using material nanogratings.

(3) The spectroscopic approach that monitors the effect of the CP potential on the shift of the atom internal energy levels.

Common challenges in all approaches include understanding and controlling the distance of the atoms from the surface (the probing distance) as well as the geometry, roughness, composition (adsorbates or parasitic fields) and chemistry of the surface in order to reduce systematic errors in the experimental measurements.

## 3.1 Mechanical approaches

### 3.1.1 Thermal beams

Thermal beams are mainly restricted to long-lived atomic states (ground or metastable states, Rydberg levels). Historically, CP forces were revealed *via* the deviation of the thermal atomic beam at razing incidence on a metallic or dielectric surface, that is using a cylinder [62], [63]. In these pioneering experiments, the atom-surface distance was not well defined. Thereafter, a series of elegant works, performed by the Yale group, have restricted the number of impact parameters with a micro-channel, defined by two parallel materiel surfaces [64], [9], [10], [65]. The transmission of a thermal sodium beam in a Rydberg state (10S to 13S) is diminished by the CP forces, which attract the atoms to the slit walls. By measuring the atomic transmission as a function of the channel width and the atom Rydberg state, the authors compared the CP forces for various atomic states [64]. The Yale group also measured in the channeled atomic beam the spectral shift of the excitation resonance toward Rydberg levels [9]. In this experiment, the authors combine mechanical effects (atomic channeling) and spectroscopic measurements. They could verify the $1/z^3$ non-retarded van der Waals regime in the micrometer scale, because the polarizability of Rydberg atoms mainly originates in atomic transitions in the mid-IR and far-IR [9], [65]. The Yale group was also able to monitor the CP interaction of a ground state sodium beam transmitted through a tunable channel and observe its retarded regime in $1/z^4$ [10].

#### *3.1.1.1 Anisotropic Casimir-Polder interactions*
All the above experimental works with atomic beams were interpreted in terms of surface-induced *scalar* forces on atoms. However, one knows that, at least in the non-retarded regime, the dipole-induced dipole atom-surface interaction has a quadrupole component, if the atomic polarizability is non-scalar. Note that in the fully retarded regime this quadrupole interaction vanishes quickly in $1/z^6$ [66], and the scalar component in $1/z^4$ stands alone. The surface quadrupole interaction induces symmetry breaking of the atomic wave functions and can couple different atomic energy levels of the same parity [67]. Quadrupole surface-induced transitions have been observed on effusive thermal beams of metastable rare gases, where the $^3P_0$ metastable state is transferred to the $^3P_2$ metastable state [52]. The energy conservation imposes a transfer from internal energy to external kinetic energy.

It modifies the atom external motion in a direction perpendicular to the surface, whereas the longitudinal velocity component along the surface is unaffected because of linear momentum conservation. Due to the large energy transfer, the angular deflection is very large, allowing a clear discrimination between deflected atoms and incident beam [52]. This transition between metastable states gives access to the very near-field surface potential at typical ranges below 5 nm. Indeed, the anisotropic part of the atom-surface interaction originates only in the electronic core ($np^5$) contribution to the metastable state polarizability (the polarizability of the excited $s$ electron being scalar). The core transitions are in the vacuum ultra-violet, implying that the anisotropic surface potential vanishes very quickly with the distance. This work has been extended to surface-induced transfer in molecular species [68], as well as to the reverse endothermic process $^3P_2$ -> $^3P_0$ [69].

Another extension has been performed analyzing surface-induced transfer between Zeeman sublevels of the $^3P_2$ metastable state [69], [70]. Indeed the quadrupole surface potential may induce transitions between Zeeman sublevels {J, M}. When a magnetic field is applied, the Zeeman energy defect is transferred to the atom motion in the same way as for fine structure coupling. Due to the smallness of the energy transfer, this work needs supersonic beams with well-defined velocity, in order to discriminate between the incident beam and deflected atoms undergoing the various M->M' sublevel transitions. Those surface-induced van der Waals-Zeeman transitions, which have been observed on supersonic beams of metastable Ne*[69], Ar*, Kr* and Xe* [70], yield to Zeeman-tunable beam splitters. For Ar* beams, a Zeeman atomic slower has been utilized to change the incident atom velocity between 560 and 170m/s. It has thus been shown that the distance range of the atom-surface impact parameter varies between 4 and 7 nm.

### 3.1.2 Cold atom experiments

The main limitation of previous works comes from the broad atomic velocity distribution with a large mean value, thus shortening the atom-surface interaction time. The development of laser cooling techniques in the late eighties has permitted the development of new CP measurements with better-defined atomic mean velocities. First works were performed by the Orsay group [13] using free falling cold ground-state Rb atoms bouncing off a dielectric surface dressed with a repulsive potential generated by a blue detuned evanescent laser field. Their reported measurements, probing atoms at distances ~50nm were in better agreement with a QED CP interaction potential but the experimental uncertainties were too large to exclude the electrostatic van der Waals model. Similar experiments were performed using magnetic films in order to create a repulsive potential close to the surface for spin-polarized atoms [71]. More recently, the repulsive evanescent optical field was used in CP measurement by the Tübingen group [14]. This time, atoms were thrown on the glass surface with a controlled, constant velocity. This was achieved by confining ultracold rubidium atoms in a combined magnetic/dipole trap before applying a sudden shift to the magnetic trap and subsequently switching it off, thus leaving the atoms free to reach the surface with a constant velocity. This work provided a measurement of the potential as a function of distance, without assuming an analytical form of the potential. The experiments showed CP retarded forces for distances between 150 and 250nm (see Figure 1).

The Tübingen group also measured the CP interaction between an atom and an ultrathin (~50nm) gold micro-grating deposited on a dielectric surface [72]. The results were interpreted using a full calculation of the complex CP potential landscape above the grating [73]. Finally, the CP potential of an isolated nanostructure was investigated by immersing a carbon nanotube in a Bose-condensed atomic cloud [74]. The CP potential of micro or nanofabricated structures finds interest in the emerging field of hybrid systems that interface solid-state platforms such as atom chips, nanofibers or photonic crystal waveguides with gold atomic clouds. The effects of the CP interactions play a critical role in many of these setups [75], [76], [77]. Further, a new generation of devices aims at utilizing CP forces in order to achieve sub-wavelength trapping (unlike optical traps, CP forces are not limited by diffraction) of cold atoms next to nanostructured surfaces [78], [79], [80], [21], [73].

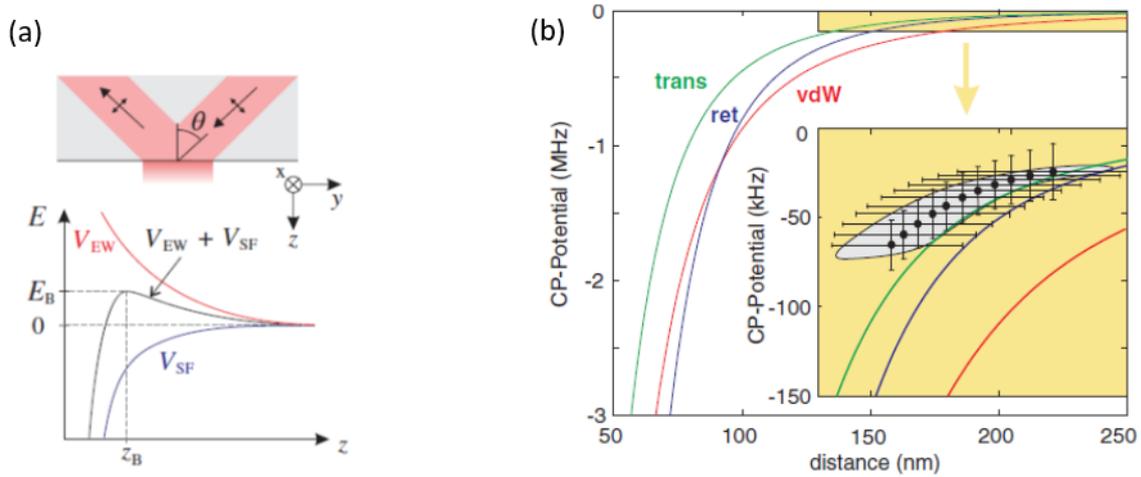

**Figure 1:** (a) Principle of the experiment performed in [14]. The combination of an attractive CP potential and a repulsive potential, created by a blue detuned evanescent field, form a barrier that reflects atoms of sufficiently low kinetic energy. (b) Experimental measurements of the CP shift as a function of distance detailed in [14] along with theoretical predictions of a purely retarded potential in $1/z^4$ (blue), purely electrostatic van der Waals potential in $1/z^3$ (red) and a full QED calculated potential (green). Reprinted with permission from H. Bender *et al*, *Phys. Rev. Lett.*, **104**, 083201, (2010) [14]. Copyright 2010 American Physical Society

Measurements of the CP force have also been performed at atom-surface separations between 5 and 7μm by the JILA group [22-27]. This distance is close to the thermal wavelength $\hbar c/k_B T \sim 7$μm and temperature effects should be considered. The first experiments were reported in ref. [15] by monitoring center-of-mass oscillations of a magnetically trapped BEC close to the surface (the theory analyzing the experiments is presented in ref. [81]). This is one of the most accurate CP measurements in the long range (see Figure 2). The measurements are in agreement with CP theory at this distance range, however thermal effects cannot be distinguished. Additionally at these distances, possible spurious systematic effects due to stray fields created by surface contaminants should be properly analysed [15]. Stray electric fields created by Rb adsorbates have been measured in independent studies performed by the same group [82], [83]. A similar experimental setup was then used to measure the far-field temperature dependence of the CP interaction out of thermal equilibrium [44]. In this work, the surface temperature was heated independently by a 1 W 860 nm laser beam and was kept at a different temperature from the surrounding vacuum. This new asymptotic regime of the CP interaction, where the atom-surface force is given by $\sim (T_S^2 - T_E^2)/z^3$, where $T_S$, $T_E$ are the surface and the environment temperature respectively, was studied theoretically in ref. [84].

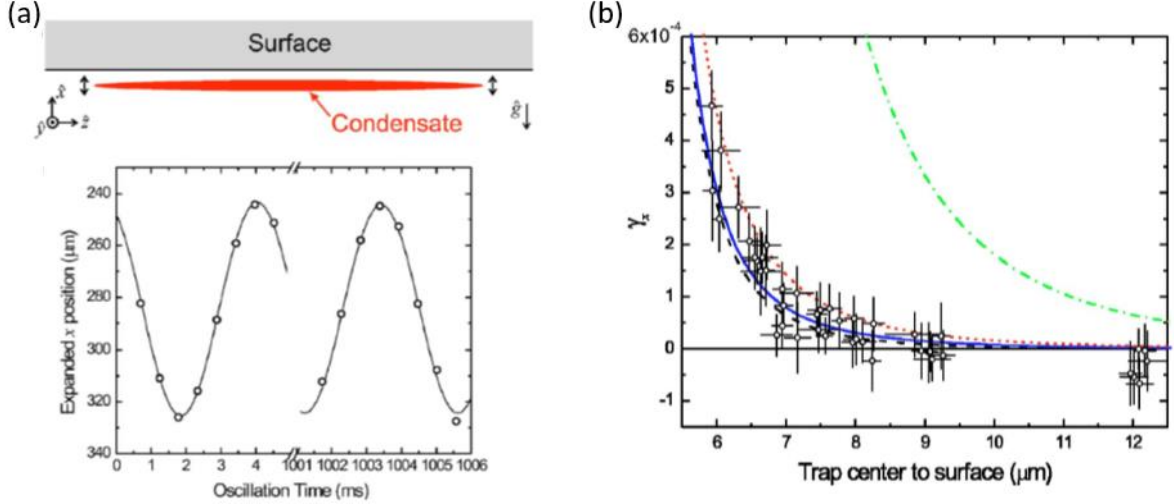

**Figure 2**: (a) A diagram to scale of the experiment reported in ref. [15] with typical data used to measure the trap dipole oscillation frequency. (b) Experimental measurement of the normalized dipole oscillation frequency $\gamma_x = \frac{\omega_x - \omega_x'}{\omega_x}$ as a function of distance from the surface. Here $\omega_x$ and $\omega_x'$ are the trap oscillation frequencies away and close to the surface (at a given distance) respectively. The black dashed, solid blue and dotted red curves represent the theoretically predicted behavior of $\gamma_x$ using the CP potential at T = 0 K, T = 300 K and T = 600 K respectively. The record sensitivity of the experiment is not enough to distinguish temperature effects at thermal equilibrium. Reprinted with permission from D. M. Harber *et al*, *Phys. Rev. A*, **72**, 033610 (2005) [15]. Copyright 2005 American Physical Society.

Further away from the surface, that is several cm away, the surface force was measured using a single excited Ba+ ion [16]. In the far-field regime of excited state particles, the interaction oscillates with a λ/2 periodicity (*i. e.,* open cavity QED) [43], in contrast to any electrostatic interactions. Therefore, ions, which are usually well-localized by strong electric traps, are good systems to study such far-field effects. Also spectroscopic works have been performed by the same group on a distant ion to extract lifetime and vacuum-field level shift [85].

One should mention that quantum reflection of ultracold atoms by a material surface needs a good knowledge of intermediate range atom-surface potentials to evaluate the reflection probability. A pioneering work was performed by Shimizu [86] who observed atom reflectivity as large as 30% for a beam of ultra-cold metastable $1s_3$ Ne* atoms falling nearly parallel to various surfaces. Theoretical fits to Ne* reflectivity measured as a function of incident velocities used an *ad hoc* CP attractive potential of the type $U \sim -C_4/(r+l_0)r^3$, with retarded and non-retarded contributions. Similar works were later performed by the MIT group [87] using a sodium Bose-Einstein condensate, confined in a harmonic trap over a silicon surface and by the group of W. Schöllkopf in Max-Planck using a helium dimmer on a grating surface [88]. More recently, the group of S. Reynaud proposed quantum reflection to tests of the weak-equivalence principle on antihydrogen [89], and quantum reflection probabilities have been predicted to exhibit discontinuities in graphene-like materials subjected to an external magnetic field [90], [91].

The group in Florence has explored an interesting experimental approach, using an optical escalator [92]. Here, an ultracold gas of strontium atoms is loaded in a one-dimensional optical lattice normal to a transparent dielectric window. The authors drag the cold cloud of atoms at the vicinity of the surface by sweeping the phase of one of the laser beam of the optical lattice. Here, one can obtain a good control of the position of the atoms with respect to the surface. An attempt to measure the CP interaction through the modification of the Bloch oscillation at few tens of micrometer was reported in ref. [92]. Unfortunately, the sensitivity of the technique was not sufficient to discriminate CP interaction with respect to earth gravitational interaction at such a large distance. Nevertheless, this optical escalator method has triggered a lot of interest to perform precise measurements of CP interaction using ultra-narrow transition [93] or matter-wave interferometric methods [94], [95] for eventually revealing non-Newtonian force at the micrometer scale [27], [95].

### 3.2 Interferometric and diffractive approaches

Interferometric and diffractive devices of the atomic wave-packet are sensitive both to mechanical effects and to surface potential. The diffractive approach has been developed in the late 1990's, initially by the Göttingen group. Here, supersonic beams of rare gases diffract on silicon-nitride (SiN) transmission nano-gratings. Experiments on rare gases (He to Kr) ground state beams [96] were followed by similar ones on He* and Ne* metastable states [97]. Also the Villetaneuse group performed similar experiments on metastable nozzle beam of He*, Ne* and Ar* [98], as well as Perrault et *al.* [99] with a supersonic sodium beam. For those fast atoms, the mechanical deflection during the transit in a nano-grating slit is negligible. Only the atom de Broglie phase shift induced by the surface potential, accumulated along a straight trajectory, has to be considered to get the single slit diffraction amplitude. The calculations were done in the non-retarded regime in $1/z^3$. Recently this approach has been extended by the Villetaneuse group to slow, mono-kinetic and velocity-tunable, Ar* beams [100] incident on a SiN nano-grating [101]. For those slow atoms (down to ~20m/s), it is necessary to take into account both the inhomogeneous CP phase shift but also the surface force which may attract metastable atoms onto the surface and subsequently eliminate them. At the reached sensitivity level, field retardation at distances as small as 50 nm is required for a full interpretation of the atomic diffraction pattern.

In parallel, full interferometric apparatuses have been devised in which a material nano-grating is inserted in one arm of a Mach-Zehnder atomic interferometer. In the Tucson group, the atom interferometer for a sodium supersonic beam was formed with three gratings [102], while the Toulouse atom interferometer for a lithium supersonic beam was made with near-resonant laser standing waves beam splitters [103], [104]. Here the experimental results are interpreted *via* non-retarded atom-surface interactions.

### 3.3 Spectroscopic approaches in vapor cells

Most of the previous methods (with few notable exceptions such as [9], [16]) are essentially sensitive to the CP interactions of ground state or metastable atoms. In contrast, spectroscopic approaches, measuring the difference of surface-induced level shifts between the probed atomic states, give access to the rich physics of short or long-lived excited atomic states. In particular, these approaches allow to probe the coupling of atoms with surface modes (analyzed in detail in section 4). In addition, spectroscopic methods also give access to surface-induced changes of the atomic state lifetime (transition linewidth), which is an integral part of the open cavity QED approach. In this section, we focus on CP spectroscopy in hot atom vapor cells that has been performed either by selective reflection on the entrance window of a macroscopic vapor cell, or transmission spectroscopy in thin vapor cells (sub-micro and nano-cells of variable thickness).

#### 3.3.1 Selective reflection spectroscopy

Selective reflection (SR) consists of monitoring the reflectivity change of a light beam incident on a window-vapor interface when one tunes the light frequency around an atomic or molecular transition. Resonant change in the vapor refraction index modifies the light reflection. Since the first demonstration of selective reflection in 1909 by Robert Wood on mercury vapor, SR has been shown to present some sub-Doppler features when performed at normal incidence on the interface and low density vapors [105], [106]. Indeed, the signal is mainly sensitive to atoms moving parallel to the interface. This sub-Doppler character can be enhanced by making use of frequency modulation (FM) on the incident probe light beam, leading to Doppler-free reflection lineshapes, as shown by the Lebedev group [107]. Since electromagnetic reflection is governed by the vapor index variations in a distance range $\sim \lambda/2\pi$ from the interface ($\lambda$ is the incident probe wavelength), FM-SR yields the atomic response close to the interface and subsequently its variations with the atom-surface potential shifting the energy levels involved in the atomic transition. More precisely, in the limit of large Doppler broadening, the FM-SR signal lineshape is proportional to the frequency-derivative of an effective electric susceptibility, χ – which takes into account the non-local response of the atomic vapor near the interface – [98] :

$$\frac{d\chi(\omega)}{d\omega} \sim \int_0^\infty dz \int_0^\infty dz' \frac{(z-z')e^{ik(z+z')}}{\mathcal{L}(\omega,z) - \mathcal{L}(\omega,z')} \qquad 3.1$$

$\mathcal{L}(\omega,z) = \frac{\Gamma}{2} - i\left(\omega - \omega_o + \frac{C_3 z^{-3}}{2}\right)$ is the Lorentzian lineshape of the bare atomic resonance, corrected by the $z^{-3}$ CP frequency shift - in the non-retarded regime -. $C_3$ is the difference between the van der Waals surface coefficients of the two energy levels of the transition, $\omega_o$ is the ~~bare~~ atomic resonance frequency, $\Gamma$ the natural transition linewidth and $k = 2\pi/\lambda$ is the wavenumber of the probe beam. In this regime, the reflection lineshape only depends on a dimensionless parameter yielding the surface-induced frequency shift, at a distance $\lambda/2\pi$, in unit of the natural linewidth. A similar formula to (3.1) can be derived in the case of retarded surface interactions [108]. FM-SR has been experimentally demonstrated by the Villetaneuse group on cesium resonance lines [11], [109], [39] and used to measure the Cs-surface interaction potential. This approach has been subsequently confirmed by many other research groups in Moscow, Garching or Boulder [110]–[113]. The interest of FM-SR spectroscopy for probing atom-surface interactions has been described in detail in a previous review article [12]. One interest of SR is to monitor the atom-surface potential in a higher excited level exhibiting large polarizability. For this purpose, stepwise excitation has been devised on the basis of an initial broadband excitation of a resonance level followed by the proper SR process. ~This has been first demonstrated on the $6S_{1/2}$-$6P_{1/2}$-$6D_{3/2}$ transition ladder of cesium, where the pumping of the $6P_{1/2}$ was guaranteed to be broadband, thanks to velocity redistribution in the resonance state via fluorescence trapping and collisional processes [114]. This makes the measurements insensitive to surface shifts on the pump transition. Selective reflection on the $6P_{1/2}$-$6D_{3/2}$ line thus gave access to the Cs $6D_{3/2}$ -surface potential which was observed to be repulsive with sapphire thanks to quasi-resonant surface phonon polariton mode [115]. This approach has inspired a number of studies on highly excited energy levels, including surface temperature dependence of the atom-surface CP potential [17] (see section 4 more details).

### 3.3.2  Thin vapor cells

Thin cell measurements started in 1996 with sub-Doppler spectroscopy [116] in micrometric cells and linear sub-Doppler spectroscopy due to Dicke narrowing [117], [118] in cells whose size goes down to a fraction of the optical wavelength (sub-micrometric size cells). For atom-surface interaction measurements, truly nanometric thickness is required. The Ashtarak Armenian team fabricated these cells [119] and measurements of the CP interaction were performed in a cell of variable thickness from 40 to 130 nm [120], [121] (see Figure 3). The advantage of thin cell spectroscopy is that one can probe vapors confined at different thicknesses in order to explore the validity of the van der Waals law $C_3/z^3$ at nanometric distances. Deviations from the van der Waals law could occur because of retardation effects [2], [7], [8] [43], [10] whose effects are predicted to be measureable even in spectroscopic experiments with low lying excited state atoms [122] even at nanometric distances. It is worth noting that spectroscopic experiments are sensitive to retardation effects on both resonant and non-resonant CP contributions.

The experiments, depicted in Figure 3, were performed on the Cs($6D_{5/2}$) level reached by $6S_{1/2}$ - $6P_{3/2}$ - $6D_{5/2}$ stepwise excitation. The experimental results were analyzed with a theoretical model based on ref. [123] taking into account the spectral shifts imposed by the two YAG windows. Ultra-thin cell experiments were also performed by the Durham group, for low lying states of cesium and rubidium [124], [125]. The experiments relied on an analysis of the wings of the laser-induced fluorescence spectra in order to extract information for atoms that move close to the walls. The authors claim that the van der Waals law is verified to within 3%. Fluorescence measurements presented in refs. [124], [125] can be complementary to thin cell transmission, but it should be noted that the experimental analysis used in ref. [124] has been put into question in ref. [126]. Additional experiments were performed by the Palaiseau group in collaboration with the Durham group [127]. This time the $C_3$ measurements were performed by transmission through a thin cell of varying thickness using the theoretical models described in [123], [120]. The $C_3$ coefficient of the D1 line was also measured with SR techniques [128].

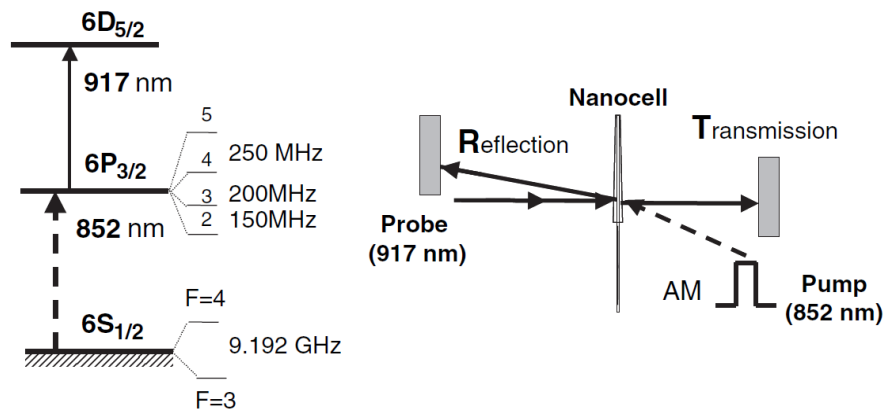

**Figure 3:** Thin cell experiments for measurements of the CP atom-surface interaction between a Cs($6D_{5/2}$) atom and YAG windows. The experiment uses a stepwise excitation scheme where cesium atoms are first pumped to the $6P_{3/2}$ level and subsequently probed at the $6P_{3/2} \rightarrow 6D_{5/2}$ transition at 917 nm. The probe laser is at almost normal incidence with respect to the thin cell walls. Both reflection and transmission of the probe are detected and analyzed in order to measure CP interactions. Reprinted with permission from M. Fichet *et al Europhys. Lett.*, **77**, 54001 (2007) [120]. Copyright 2007 IOP Publishing.

# 4. Tunable Casimir-Polder interactions

Surface plasmon polaritons or phonon polaritons have a profound effect on near-field Casimir-Polder (CP) interactions that strongly depend on the evanescent modes of the vacuum field [129]. Therefore, the coupling of atoms to surface polariton modes allows to tune the strength of CP interaction in the non-retarded regime [115]. In particular, temperature provides an important knob to tune CP strength [17] because thermal population of the evanescent surface polariton modes creates quasi-monochromatic thermal fields, whose spectral width mainly depends on the Q-factor of the surface resonance [18], [19]. This is in striking contrast to far-field blackbody radiation, which is dominated by propagating modes, and thus displays a broad frequency spectrum. Another effective way to tune the CP interactions is by nano-structuring the surface. The booming field of nanostructures and metamaterials now gives us the possibility to go beyond optical materials that exist in nature [130] and allows us to engineer evanescent modes, thus changing the fabric of the quantum vacuum [131]–[134]. This field is still new, and although there are still challenges that lie ahead, it offers great potential for subwavelength trapping of cold atoms by use of CP forces [135] and for quantum technology applications.

In this section, we first give an overview of the works investigating the coupling of atoms to surface resonances. We then give a brief theoretical description of thermal effects in the near field of the CP interaction [35], [32], which are due to the thermal excitation of surface polariton modes. Such effects were experimentally demonstrated with selective reflection spectroscopy in atomic vapor cells [17], [136]. Finally, we describe experiments reporting tunable CP interactions by resonantly coupling atoms to metamaterials with surface plasmon resonances [131], [132].

### 4.1 Coupling of atoms with surface resonances

The near-field coupling of atoms with surface-polaritons has been theoretically investigated in [129], using the linear-response theory developed by Wylie and Sipe in refs. [7], [8]. These theoretical works make the assumption that thermal energies are always much smaller than transition energies and are, therefore, limited to zero temperature ($T = 0$). The first experimental demonstration of atom-polariton coupling was done with stepwise selective reflection spectroscopy by Failache et al [115] using Cs($6D_{3/2}$) atoms. This atomic level presents a dipole transition at 12.14 μm ($6D_{3/2} \rightarrow 7P_{1/2}$) that is strongly coupled to a sapphire surface resonance at ~12μm. The experiments are performed at temperatures where the thermal excitation of the sapphire polariton modes is negligible, so they can be considered as $T = 0$ experiments. It is worth mentioning that although Failache et al [115] demonstrated CP repulsion, the measurement of the $C_3$ coefficient was not in complete agreement with theoretical predictions, thus underlying the sensitivity of CP experiments to the dielectric properties of the surface [137] and the need for more accurate measurements of the dielectric constant of materials.

A general theoretical treatment of resonant atom-surface interactions as function of temperature was performed by ref. [35] with emphasis on calculations for atoms passing through cavities. Experiments with Rydberg atoms passing through cavities were indeed performed in the early 90s [9], [10], but did not demonstrate near-field temperature effects on the atom-surface interaction [138]. Instead, they focused on the impact of blackbody radiation inside a cavity [139]. The treatment of refs. [7], [8] was extended to a finite temperature environment in ref. [32]. Although the treatment of ref. [32] is valid at all distances, the main focus is on near-field atom-polariton coupling in selective reflection experiments. A general theoretical treatment of temperature dependent CP interactions is also presented in ref. [36].

Further theoretical works, focusing on near-field CP interactions [45], [140] with thermally excited surfaces, demonstrate that the temperature dependence of atom-surface interactions is due to the thermal excitation of surface modes (near-field thermal emission) and that CP interactions can be largely independent of temperature when atomic transitions are far detuned from polariton resonances. In fact, near-field thermal emission, dominated by evanescent modes, is largely monochromatic at the polariton frequency $\omega_S$ [18], [19]. Therefore, the near-field CP dependence on temperature can be considered as a dipole force exerted on the atoms by the thermal fields at the vicinity of the hot surface. Atoms are either repelled or attracted toward the surface depending on the detuning (red or blue) of the evanescent thermal fields compared to the atomic dipole coupling frequency $\omega_{ab}$. For this reason CP shifts for Rydberg atoms [141] or even molecules [140] can remain temperature independent, even though they present strong couplings (especially in the case of Rydberg atoms) at low energies.

## 4.2 Near-field thermal effects in the Casimir-Polder interaction

The dielectric constant of dielectric media is often described as a sum of Lorentzian resonances. This is because the response of a dielectric to an external driving field can be approximated to an externally driven harmonic oscillator problem. For example, if only one isolated volume resonance exists in the frequency range of interest, one can write the material dielectric constant as follows:

$$\varepsilon(\omega) = \varepsilon_\infty + \frac{(\varepsilon_o - \varepsilon_\infty)\omega_V^2}{\omega_V^2 - \omega^2 - i\omega\Gamma} \quad (4.1)$$

Here $\varepsilon_\infty$, and $\varepsilon_o$ are the limiting values of the dielectric constant at high or low frequencies respectively, $\omega_V$ is the volume resonance frequency and $\Gamma$ the dissipation parameter. The above model and some variations, assuming that the dissipation constant is itself a function of frequency $\Gamma(\omega)$ (accounting for lattice anharmonicity), have found extensive use in the study of dielectric properties of materials [142], [143], [144], [145]. Here we use the assumption of a single resonance, in Eq. (4.1), that allows the derivation of analytical formulas making the physics of thermal CP interactions more transparent.

The surface response of the dielectric with an infinite planar interface is given by the expression:

$$S(\omega) = \frac{\varepsilon(\omega) - 1}{\varepsilon(\omega) + 1} \quad (4.2)$$

Thus, the surface response is also a Lorentzian function written as [45]:

$$S(\omega) = S_\infty + \frac{(S_o - S_\infty)\omega_S^2}{\omega_S^2 - \omega^2 - i\omega\Gamma(\omega)} \quad (4.3)$$

$S_o$, $S_\infty$ give the surface response at the extrema of the spectrum, linked to $\varepsilon_\infty$, and $\varepsilon_o$ via Eq. (4.2). The surface resonance $\omega_S$ is related to the volume resonance by the following expression:

$$\omega_S = \sqrt{\frac{\varepsilon_o + 1}{\varepsilon_\infty + 1}} \omega_V \quad (4.4)$$

Using Eq. (4.3), the Matsubara summation (described in section 2) can be calculated analytically [45]. For a two-level transition $|a\rangle \to |b\rangle$ at frequency $\omega_{ab}$, the temperature dependent part of the near–field free energy shift of level $|a\rangle$, $\delta F_a^T = \delta F_a(T) - \delta F_a(T = 0)$, can be simplified to:

$$\delta F_a^T = -\frac{C_3^{a \to b}}{z^3} \left\{ 2n(\omega_S, T) \frac{\omega_{ab}\omega_S(S_o - S_\infty)\left[(\omega_{ab}^2 - \omega_S^2) + \frac{\Gamma^2}{2}\right]}{(\omega_{ab}^2 - \omega_S^2)^2 + \Gamma^2\omega_S^2} \right\} \quad (4.6)$$

where $\omega_{ab}$ is positive or negative for an upwards (absorption) or downwards (emission) coupling respectively. $n(\omega_S, T)$ is the average photon occupation for Bose-Einstein statistics (see Section 2) at the surface polariton frequency $\omega_S$ and not the atomic transition frequency $\omega_{ab}$, demonstrating that the near-field temperature dependence of the CP interaction depends on the thermal excitation of surface polariton modes only. For an absorption coupling, the energy shift is negative, pulling the atoms toward the surface, when the thermal field frequency ($\omega_S$) is red detuned with respect to the atomic transition ($\omega_{ab}$), whereas for a blue detuned thermal field the atom is repelled from the surface. The opposite scenario occurs when the coupling is in emission. It is worth stressing that the CP tunability offered by atom-polariton coupling is a near-field effect and is washed out in the far-field of CP interaction [17], [45] when retardation effects become important.

### 4.2.1 Thermal Casimir-Polder interaction: Experiments

The above theoretical considerations show that the near-field thermal effects allow for CP tunability. This is a distinct difference from the temperature dependent CP interaction exerted on ground state atoms. In this case, the temperature dependence becomes predominant essentially in the far-field regime [6], [15], [44], where the interaction is simply proportional to temperature $\delta F_a(T) \propto -\frac{T}{z^3}$, at distances longer than the thermal wavelength (at thermal equilibrium). Such temperature effects remain without experimental demonstration, although significant efforts were made toward this direction by the JILA group [15]. Temperature effects should also be taken into consideration for precision CP measurements with ground state atoms, as discussed for example in refs. [93] [146] (see Section 3 for more details).

Near-field thermal effects in the CP interaction have been studied exclusively with spectroscopic methods inside vapor cells [17], [143], [137]. A key element of these experiments is the probing of excited alkali atoms presenting many allowed dipole transition that can be coupled to surface polaritons. A significant amount of work toward identifying the surface polariton resonances of dielectric material was done in ref. [147]. Nevertheless, the choice of dielectric materials that can be used for vapor cell fabrication is limited mainly because high temperature operation is required under the presence of chemically aggressive alkali vapors. The first experiments aiming at measuring the near-field temperature dependence of the CP interaction were performed with a calcium fluoride surface ($CaF_2$). The interest of fluoride dielectrics, such as $CaF_2$ and $BaF_2$ is their isolated surface resonances at 24 and 35 μm respectively that can be coupled with the strong dipole transitions from the $8P_{3/2}$ level, $8P_{3/2} \to 7D_{3/2}$ at 36.09 μm [143], [136]. The coupling of the $8P_{3/2} \to 7D_{3/2}$ transition to the blue detuned $CaF_2$ polariton leads to a reduction of the van der Waals coefficient with temperature. Exact theoretical predictions of the $Cs(8P_{3/2})$-$CaF_2$ interaction as a function of temperature were shown in ref. [143] using dedicated measurements of the $CaF_2$ dielectric constant. In order to probe the interaction of cesium atoms with a $CaF_2$ surface, an original cell design was proposed and tested in ref. [128] by performing selective reflection measurements on the D1 transition of cesium. Selective reflection measurements, of the CP interaction between $Cs(8P_{3/2})$ and a $CaF_2$ surface were reported in ref. [136]. Nevertheless, these measurements did not show a temperature dependence of the van der Waals coefficient in contradiction to the theoretical predictions. This disagreement was attributed to cesium chemical attacks on the $CaF_2$ tube, whose chemical integrity seemed to be seriously compromised (the cesium vapor infiltrated inside the volume of the $CaF_2$ tube) [136].

The near-field temperature dependence of the CP interaction at thermal equilibrium was demonstrated for the first time in a vapor cell experiment measuring the interaction between a Cs($7D_{3/2}$) atom and a sapphire surface [17]. In this experiment, the temperature dependence comes from the coupling between the sapphire polariton at ~12 μm with the cesium upward $7D_{3/2} \rightarrow 5F_{5/2}$ transition at 10.83 μm. Due to the red detuning of the near field thermal emission, the $C_3$ van der Waals coefficient is predicted to increase with temperature (see Figure 4). Experimental results and theoretical predictions are in a quite remarkable agreement for temperatures up to 1000 K. This could be because the sapphire surface polariton is relatively far detuned with respect to the atomic resonance, thus making the theoretical predictions less sensitive to the details of the sapphire dielectric constant [17]. Perhaps equally critical to the success of the experiment was the use of a sapphire window that has been shown to be robust and durable in the presence of high temperature cesium vapors [148]. The sapphire cell used for this experiment is described in detail in ref. [17], [149].

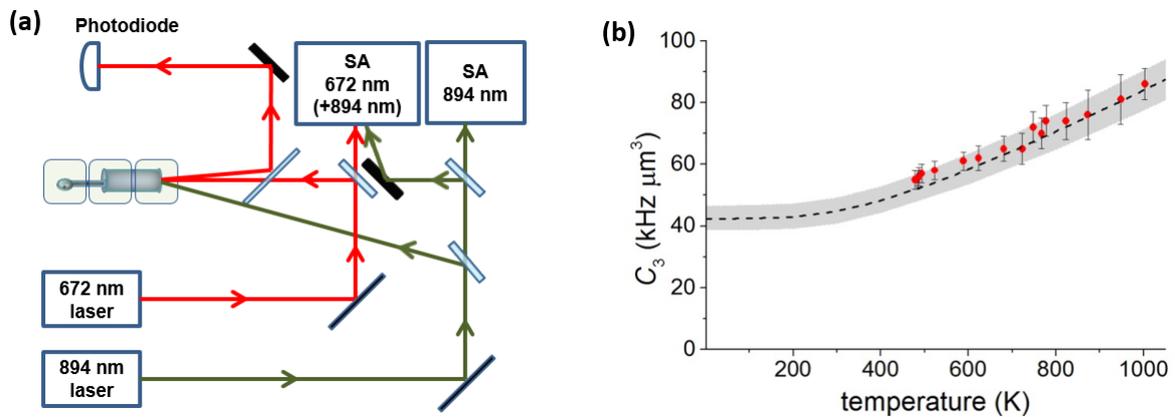

**Figure 4:** (a) Experimental set-up of the experiments reported in ref. [17]. The experiments use a stepwise excitation scheme, pumping the atoms to the $6P_{1/2}$ level and then performing selective reflection on the $6P_{1/2} \rightarrow 7D_{3/2}$ level. Collisions and radiation trapping create a thermal population of excited atoms at the $6P_{1/2}$ level. (b) Experimental measurement of the van der Waals coefficient of the Cs($7D_{3/2}$)-sapphire interaction as a function of temperature, along with theoretical predictions based on different measurements of the dielectric constant of sapphire (dotted and dashed black lines). Reprinted with permission from A. Laliotis *et al, Nat. Commun.*, **5**, 5364, (2014) [17]. Copyright 2004 Springer Nature.

### 4.3 Tunable Casimir-Polder interaction using metasurfaces

In the section 4.2, we discussed the near-field temperature dependence of CP interactions. Thermal effects can indeed be used to tune the CP interaction of excited atoms with dielectric materials. This is achieved by modifying the thermal population of surface modes, proportional to the average number of photons $n(\omega_S, T)$ governed by Bose-Einstein statistics. According to Eq. 4.6 tunability is essentially limited to dipole couplings whose frequency is near resonant with surface polariton resonances. In addition, achieving tunability for experimentally realistic temperatures requires surface phonon polariton modes whose frequency lies in the mid or far infrared (with sapphire being the best candidate). The above constraints suggest that tuning CP by temperature is essentially limited to high-lying atomic states that present allowed dipole transitions at mid and far-infrared wavelengths.

In order to address low-lying states whose allowed dipole couplings lie in the near-infrared and visible range of the spectrum a more radical approach has been demonstrated, involving the use of metamaterial (nanostructured metallic surfaces). Engineering the geometry of the metamaterials allows to tune the dielectric properties and the surface-plasmon resonances of the metamaterial as depicted in Eq. (4.4) [132]–[134].

### 4.3.1 Electric dipole coupling

CP interaction was revealed using selective reflection spectroscopy and a nanofabricated surface, immersed in an atomic vapor, as depicted in ref. [131] and in Figure 5 (left panel). Here, nano-slits on a 50 nm thick silver film were engraved. The geometry of the nano-slits, that is length and width, were adjusted such that the surface-plasmon resonance frequency is scanned around the D$_2$-line resonance of Cesium atoms at 852 nm. Tunable CP interaction was experimentally observed as discussed in ref. [132]. The results are summarized in the right panel of Figure 5A and 5B, where the real and imaginary part of differential van der Waals coefficient $\Delta C_3 = C_3^{(e)} - C_3^{(g)}$ is plotted as function of the metasurfaces plasmonic resonance wavelength. $C_3^{(g)}$ and $C_3^{(e)}$ are the van der Waals coefficients of the ground and excited state, respectively. A resonance profile of $\Delta C_3$ is observed, having a linewidth of the order of the plasmonic resonance linewidth. The main contribution at the origin of the modification of $\Delta C_3$ comes from the resonant CP term [11] when a photon is virtually emitted from the excited state of the atom into the metamaterial before being reabsorbed by the atom. Hence, the surface-plasmon mainly modifies the $C_3^{(e)}$ term. When plasmonic and atomic resonances are coinciding, the emitted photon might be effectively transferred to the surface-plasmon leading to a non-zero imaginary term for $\Delta C_3$. The atomic lifetime decreases with the presence of the surface plasmon resonance. This transfer process is irreversible because of ohmic losses in the silver metamaterial. The experimental data in Figure 5A and 5B (right panel) are compared to theoretical predictions in Figure 5C and 5D without retardation effect (dotted-dashed line), and with retardation effect (shaded gray surface). We note that the predictions are weaker than the experimental values, indicating that some phenomena are not captured in the theoretical model.

### 4.3.2 Electric quadrupole coupling

For a Gaussian beam, incident on the metasurface, its transmission can be decomposed into far-field and near-field contributions. The far-field contribution comes from the modification of the index of refraction of the nano-structured film, which acts as a Fabry-Pérot etalon. The near-field contribution is characterized by an evanescent wave, with a typical spatial damping of the optical field linked to the period of the metasurface. The contribution of both effects was investigated using the $6^2S_{1/2} \rightarrow 5^2D_{5/2}$ quadrupole transition of cesium at 685 nm [133]. Indeed, quadrupole transitions are expected to be enhanced by a strong electric field gradient [150]. Such an enhancement was not observed due to the component of the atoms thermal velocity parallel to the metasurface, which averages to zero the evanescent wave contribution, because of transit-time broadening. Thus, only the far-field etalon effect is at the origin of the modification of the resonant quadrupole CP interaction using a thermal atomic vapor. Two strategies can be considered to observe the near-field enhancement. The first one consists of using a quantum emitter at rest, by suppressing the Doppler effect using either laser cooled atoms or artificial atoms like crystal defect [151]–[153]. The second method consists of using isotropic translation-invariant 2D materials [154]. Here, the coupling is expected to be insensitive to the parallel components of the atom velocity, thus suitable for thermal atomic vapor.

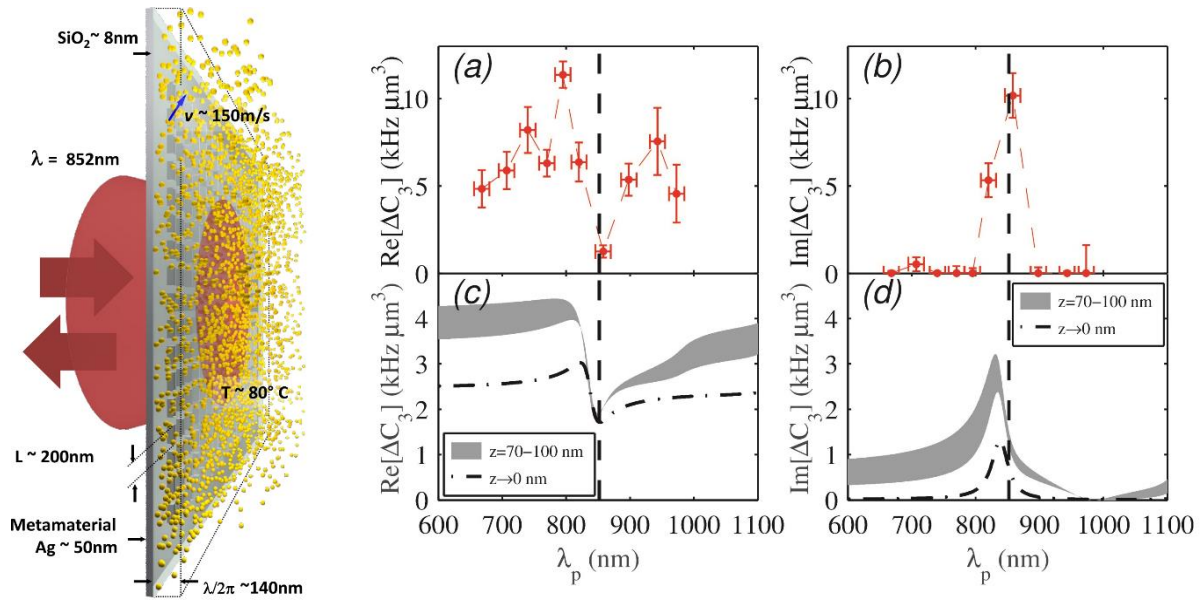

**Figure 5:** Left panel: Schematic of selective reflection spectroscopy set-up on a cesium vapor (artistically represented as yellow dots) with a metasurface made of a periodic arrangement of nano-slits. Reprinted with permission from Aljunid *et al*, *Nano Lett.*, **16**, 3137, (2016) [131] (copyright@2016, American Chemical Society). Right panel: Real part (a) and imaginary part (b) of the differential van der Waals coefficient (red dots) as function of the plasmonic resonance position. The panel (c) and (d) correspond to the theoretical prediction in the non-retarded regime (dashed-dotted curve) and in the retarded regime (shaded gray surface). Reprinted with permission from Chan *et al*, *Sci. Adv.* **4**, aao4223 (2018) [132]. Copyright 2018 AAAS.

# 5. New materials

The recent development of nanoscale fabrication techniques, such as e-beam milling, deep UV photon lithography, chemical vapor deposition, molecular-beam epitaxy, and exfoliation of 2D materials [155] has opened avenues to create new material for optics [130], [156], [157]. In particular, the confinement of electrons in 1D or 2D leads to materials with strong electric permittivity $\varepsilon$ or magnetic permeability $\mu$ anisotropy. If surface-plasmon-polariton (SPP) resonances are present, some components of the $\varepsilon$ or $\mu$ tensors might become negative leading to left-handled or hyperbolic metamaterials which have triggered a lot of interest due to their super-resolution properties for imaging application (see refs. [158], [159] for a review).

### 5.1 Coupling to surface-plasmon-polariton resonances

Placing a quantum emitter (QE) in the near-field of a (meta)material with SPP resonances, will introduce new excitation and decay channels that deeply modify the QE emission properties. For instance it has been demonstrated that coupling atomic resonances to SPP resonances leads to Fano-like excitation profile using a metallic layer [160], [161] or a metamaterial [131], [134]. In those experiments, the SPP-QE coupling strength remains moderated because of the small Q-factor and weak confinement of the SPP mode. Improvement on both latter aspects might lead to light-matter strong coupling regime where coherent evolution prevails over spontaneous emission (see refs. [28], [162], [163] for review). For example, strong-coupling has been reported between a dye molecule [164] or a quantum dot [165] and a plasmonic nanocavity at room temperature. Rivera and co-authors also pointed out that strong plasmonic confinement reduces the effective size of the electromagnetic field well below its vacuum wavelength, opening high order electric and magnetic multipole excitations [154].

Cooperative emission of a QE ensemble might also be enhanced by the presence of SPP resonance. This point was experimentally addressed by Stehle, Zimmermann and Slama studying the emission properties of an ultracold gas of rubidium atoms at the vicinity of a metallic layer [161]. The coupling of two atoms mediated by a material was also addressed theoretically and an enhancement or reduction of the atom-atom coupling was shown depending on the choice of the material and the orientation of the atomic electric dipoles [166]–[168], as well as long range dipole-dipole interactions [169].

### 5.2 Search of Casimir-Polder repulsion

The development of nanotechnologies goes hand in hand with drastic reduction of the system size. Under this perspective, understanding Casimir or CP interaction and even engineering it with new materials, becomes a major challenge for micro and nano devices [170]. Ultimately, if a composition of one or several materials is able to reverse the sign of the force going from attractive to repulsive interaction, one can suppress friction opening the door for (opto-)mechanical nanodevices. An important step was achieved in 2011 with the prediction of a repulsive Casimir force between a pair of axionic topological insulators (without dispersion in the axionic coupling strength) [171], and subsequently, a similar prediction of Casimir repulsion between a pair of Chern insulators at zero temperature [172]. The repulsion arises at sufficiently large separations, where the Casimir-Lifshitz energy is dominated by the low-frequency modes, for which the longitudinal conductivity is negligible and the Hall conductivity is topologically quantized. As the sign of the Hall conductivity of a Chern insulator can be changed by simply turning the Chern insulator around, this also enables one to control the sign of the Casimir force.

Achieving CP repulsion is also of considerable practical importance to develop for instance nano-sensors or robust QE-material interface for quantum technology applications. We discussed in Section 4, how CP interaction can be tuned and eventually reversed by a resonant coupling between a QE and surface modes. However, these examples involve a QE at an excited state with short lifetime, thus considerable reducing its interest. Would it be possible to obtain CP repulsion for long-lived state of QE? To the best of our knowledge, no experimental demonstration of such inversion of the CP interaction, which necessarily should occur over a broad frequency range, has been reported so far. Nevertheless, numerous theoretical proposals, involving such new material as nanobodies, nanostructures, magnetic and topological materials, have predicted CP repulsion, which should inspire future experimental realizations. We give now an overview of the various systems that are predicted to give CP repulsion

### 5.2.1 Nanostructures

Repulsive CP interaction can be obtained using several geometrical configurations for an atom in a long-lived state. It was shown for example that a dielectric slab with a circular opening creates repulsive CP interaction for an atom with anisotropic polarization [173], [174]. Anisotropy in polarization of the QE is a key ingredient in the prediction of CP repulsion. Such a requirement is also pointed out in other theoretical works as: a conducting toroid [175] and a 1D nano-grating [176], in the non-retarded regime; a metallic plate with a hole [177], [178] (see Figure 6); a dielectric ring [179]; near edges [180] and at the wedge of a conductor [173]. To better understand the impact of nano-structuration of a macroscopic body on the CP interaction, P. Venkataram and co-authors have derived the lower (attractive) and upper (repulsive) bonds on the CP force [181]. They found that nano-structuring does not drastically improve the attractive character of the CP force with respect to planar materials, but repulsion can be largely improved with respect to the current designs. This work opens the door for search of optimum nano-structuring of material for CP repulsion.

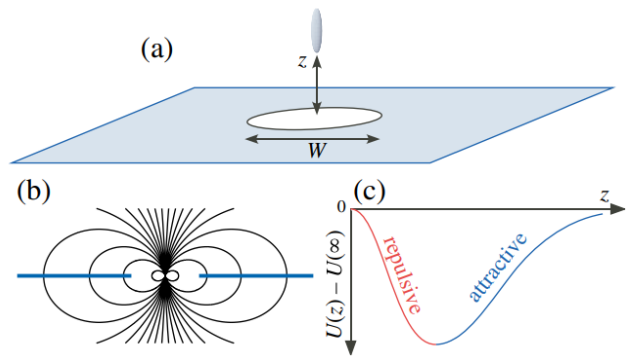

**Figure 6**: (a) Schematic representation of an elongated metal particle (polarization along z) and a thin metal plate with a hole. (b) At $z = 0$, the fluctuating dipole are unaffected by the plate. (c) Schematic of the particle-plate interaction energy. It is zero at $z = 0$ and $z \to \infty$, and attractive for $z \gg W$, so a repulsive region should be present close to the plate. Reprinted with permission from M. Levin *et al, Phys. Rev. Lett.*, **105**, 090403, (2010) [177]. Copyright 2010 American Physical Society.

Nanostructuration has also been used to create left-handed materials. From perfect reflecting left-handed material, repulsive CP force was predicted [182]. This effect comes from the extra $\pi$-phase shift at reflexion due to the negative refractive index. However due to Kramers-Kronig relations perfect lensing could not be maintained at all frequencies [173], reducing repulsive CP force to short-lived excited state [183], where resonant transitions to low-lying states largely contribute to the CP interaction [8], [115].

### 5.2.2 Magnetic polarization and magnetic materials

If one considers a magnetically polarized atom, repulsive CP interactions are found with conducting walls [184], and for variety of slabs such as perfect conducting, dielectric, and superconducting [185]. It was also shown that the general attractive CP interaction of electric dipole can be counteracted by a repulsive magnetic interaction created by a permanent magnetic moment [186]. The latter can be dominant providing that the permanent magnetic moment is large enough [187]. If a magnetically polarized atom is found to have repulsive CP interaction in front of an electrically polarized wall, it was shown that the opposite configuration, that is an electric dipole in front of a magnetic material, also leads to repulsive CP interaction [188]. One shall note that this configuration does not obey the same asymptotic power laws than the standard attractive CP interaction, see Table 1. With the development of metamaterial showing strong magnetic resonance in the microwave [189] , THz [190] and IR [191] spectral region, interest for magnetic materials to generate repulsive CP interaction has recently grow [27], [183], [188], [192], [193]. However, experimental evidence of such repulsion of magnetic material for long-lived atomic states is still lacking. Indeed, strong permeability in actual metamaterials is achieved using resonant nanostructure (for example; split ring resonator) which limits the interesting magnetic property to a narrow frequency range. It is then challenging to overcome the broad-spectrum attractive electric response of the material.

| distance | short | | long | |
|---|---|---|---|---|
| polarizability | e ↔ e | e ↔ m | e ↔ e | e ↔ m |
| atom ↔ half space | $-\dfrac{1}{z^4}$ | $+\dfrac{1}{z^2}$ | $-\dfrac{1}{z^5}$ | $+\dfrac{1}{z^5}$ |
| atom ↔ thin plate | $-\dfrac{1}{z^5}$ | $+\dfrac{1}{z^3}$ | $-\dfrac{1}{z^6}$ | $+\dfrac{1}{z^6}$ |
| atom ↔ atom | $-\dfrac{1}{z^7}$ | $+\dfrac{1}{z^5}$ | $-\dfrac{1}{z^8}$ | $+\dfrac{1}{z^8}$ |
| half space ↔ half space | $-\dfrac{1}{z^3}$ | $+\dfrac{1}{z}$ | $-\dfrac{1}{z^4}$ | $+\dfrac{1}{z^4}$ |

**Table 1**: Asymptotic power laws of the force between various polarizable object. The heading e and m stand for electric and magnetic interaction. Reprinted with permission from S. Y. Buhmann *et al, Phys. Rev. A*, **72**, 032112, (2005) [188]. Copyright 2005 American Physical Society.

### 5.2.3 Topological materials

Topological materials are materials whose band structures have nontrivial topology, i.e., there are Dirac points which are effectively topological charges (magnetic monopoles) in momentum space, and are thus characterized by nonzero Chern numbers which are (positive or negative) integers [194]. Unconventional charge and spin transports results from non-trivial topology such as dissipation-less surface spin-preserved conduction in topological insulator, and integer, spin and anomalous quantum

Hall effects [195]. The unusual conductivity of topological materials naturally leads to interesting electromagnetic and optical properties [196]. In particular, it was shown that electromagnetic effects of a magnetic topological insulator can be described by an effective axion term [197], [198]. The optical properties of such axionic system was further investigated by Crosse and co-authors [199]. They show that the discontinuity of the axion at a topological insulator's interface with the vacuum causes the transverse electric and magnetic component to mix, a property closely related to the Hall conductance of an anomalous quantum Hall material, which breaks the time-reversal symmetry and leads to nonreciprocity. Using these results, the same authors derive the CP interaction of a perfectly reflecting nonreciprocal medium with a nondispersive axion coupling strength, and show that oscillation between repulsive and attractive CP interaction can be achieved under specific orientations of the atomic dipole [200], see Figure 7. Those results were recently confirmed and extended to more realistic nonreciprocal materials, i.e., Chern insulators with dispersive conductivity response described by the Qi-Wu-Zhang model [201]. Similar mechanisms are also present in graphene subjected to external magnetic field, indicating that an anomalous quantum Hall effect is responsible for nonreciprocity and discontinuities of the CP potential [202]. The idea of constructing an atom trap using a cavity made of an internal layer of negative refractive-index metamaterial and an external layer of topological insulator has also been proposed [203]. Using an effective axion term, Marachevsky and Pis'mak show that an atom in front of a Chern insulator can be described by an antisymmetric part of the atomic dipole moment correlation function, which is a key ingredient for CP repulsion [204]. Other examples of such antisymmetric components were used in anisotropic quantum vacuum environment to predict quantum interference in V-type transitions [205] and long-lifetime ground state coherence for Λ–type transitions [206].

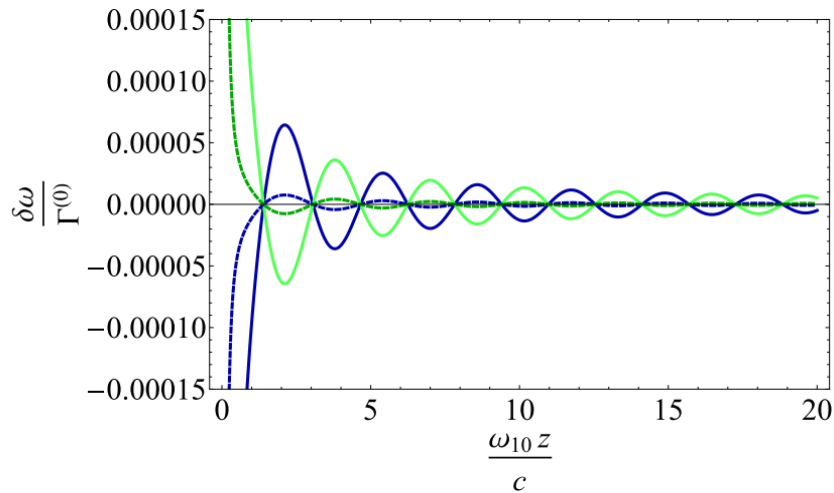

**Figure 7**: Frequency shift of a two-level atomic dipole normalized to the vacuum transition linewidth as function of the surface distance in wavenumber unit. The plain (dashed) lines correspond to a material with (without) axion contribution. The green and blue curves correspond to a positive and negative axion coupling. Reprinted with permission from S. Fuchs *et al*, *Phys. Rev. A*, **95**, 023805, (2017) [200]. Copyright 2017 American Physical Society.

### 5.3 Giant thermal Casimir-Polder interactions with 2D crystals

Last but not least, we look at 2D crystals, in particular graphene; these are materials with atom-sized thicknesses (∼ 0.1 nm), and thus not quite the same as nanoscale metamaterials. As we mentioned in Section 2, in the Lifshitz regime, the power-law scaling of $z^{-3}$ in the CP energy originates in the zero-frequency mode, which come from photons for a system consisting of regular dielectric materials bathed in radiation. Photons propagate with speed $c$, and thus dimensional analysis dictates that the thermal wavelength must involve the ratio $\hbar c/k_B T$. On the other hand, it was pointed out in ref. [207] that the thermal wavelength is shorter for the case of graphene, for which there is an additional zero-frequency mode that comes from thermally excited charges, owing to the absence of graphene's gap. The charge fluctuations propagate with Fermi velocity $v_F$ (of the order of $10^6$ ms$^{-1}$), which implies that the corresponding thermal wavelength is $\lambda_T = \hbar v_F/k_B T$. Hence, the Lifshitz regime can occur at the much smaller separation of 150 nm rather than 7 µm. There can thus be a giant thermal effect for graphene, where the thermal Casimir force is much larger. To date, one experimental proposal has been made to measure this effect [208]. The impact of the thermal effect on the atom-surface interaction is theoretically investigated in [209], for a ground-state atom interacting with graphene on four different kinds of substrate: gold, silicon, aluminum oxide and silicon dioxide. It was funded that thermal effect becomes stronger and sets in at shorter separations with a decrease in the substrate's static dielectric permittivity, being strongest for graphene-coated silicon dioxide.

# 6. Outlook

In this review, we give a quick overview of the quantum electrodynamics (QED) framework that describes atom-surface interactions, pioneered by Casimir and Polder. We also describe a series of experiments that validate the QED approach from nano to macro scales, including the effects of vacuum and thermal fluctuations as well as surface nanostructuring. After such thorough experimental scrutiny, it seems only natural that the study of atoms or molecules in interaction with dielectric/metallic surfaces is now referred to simply as Casimir-Polder (CP) physics.

The revolutionary concept of exploiting CP interactions, beyond simply studying them, has recently emerged in the growing field of quantum technologies. In particular, utilizing CP forces to achieve atom trapping at subwavelength distances from dielectric walls, without diffraction limitations, has attracted significant theoretical and experimental interest [21]. Nevertheless, one cannot achieve stable 3D trapping of atoms relying exclusively on attractive CP forces (no-go theorem) [210]. For this reason, tuning atom-surface interactions is a key issue in CP physics. Arguably, the biggest achievements toward CP tuning have been demonstrated with excited state atoms taking advantage of the resonant coupling that occurs between atomic dipole transitions and surface polariton resonances [129]. After the pioneering demonstration of CP repulsion in atomic vapor cells [115], the possibility of tunable CP interactions via temperature was also demonstrated [17]. A more recent breakthrough in CP tunability was the use of metamaterials with tunable surface plasmon polariton resonances [132]. All the above schemes are severely limited by the short lifetime of the excited states, but it has been proposed that dressing ground (attractive) states with excited (repulsive) states with an optical field can be a way to contour such limitations [135], toward all-CP trapping. Over the last decade, several important theoretical works have also proposed other schemes to achieve CP repulsion involving either nano-structure, magnetic or topological material. Anisotropy in the electromagnetic field coupling for both the quantum emitter and the material seems to be the key ingredient in the prediction of CP repulsion.

An interesting direction in the field of CP experiments is highly excited (Rydberg) atoms. The magnified properties of Rydberg atoms (such as atomic radius, polarizability, and dipole moment) make them interact strongly with their neighbors thus favoring the observation of collective phenomena and making them good candidates for quantum technology applications. Nevertheless, Rydberg atoms also interact strongly with surrounding surfaces (the interactions scale as $n^4$, where n is the effective quantum number). Eventually, many common approximations made in the second-order perturbation QED theory break down in the case of Rydberg-surface interaction. For instance, the dipole-dipole interaction approximation is no longer valid when the atomic radius is non negligible compared to the atom-surface distance and quadrupole-quadrupole interaction terms need to be considered [141]. In addition, perturbative approaches are not applicable when the CP shifts are not negligible compared to the energy level spacing [211]. It is probably for this reason that experimental attempts to probe CP interactions with high-lying Rydberg atoms in thin cells [212] or porous media [213], [214] have been in disagreement with standard CP theory. Another fascinating aspect of CP interactions with Rydberg atoms is the coupling with surface phonon-polariton resonances. Although most Rydberg couplings occur in the microwave range, far from the surface resonances of dielectrics (mid-infrared range), the possibility of achieving strong coupling of Rydberg atoms with resonators and structured surfaces has been theoretically investigated [215]. Such schemes could provide a path for tuning Rydberg-surface interactions with possible implications in the field of quantum technologies interfacing Rydberg atoms with solid-state platforms such as thin cells [216], hollow core fibers [217] and nanofibers [218].

Mastering CP interactions is also of key importance for fundamental test in physics. For instance, accurate measurement of the CP interaction would allow searching for possible deviation of the gravitational interaction or the presence of a fifth fundamental interaction at the micrometer length scale or lower. However, measurements reported so far, have limited relative accuracy, around 10%, and limited control of the atom-surface distance. In addition, in many cases, interpreting the experimental results requires an *a priori* assumption on the distance dependence of the interaction that is usually validated by fitting to the experimental data. A direct and accurate measurement of the CP interaction will require measurement of the energy shift of quantum emitters located at a known distance from the interface. A possible approach was tested by Sorrentino et *al.*, where the authors transport a cold gas close to a transparent glass with a moving optical lattice [92]. A similar experiment aiming at measuring CP using Raman interferometry of Wannier-Stark states has been proposed by the Paris's group at SYRTE [146], [219]. Ultra-narrow transitions in two-electron atoms have been also considered for precise spectroscopic measurements [93].

Finally, another important experimental goal of CP physics is precision measurements in molecule-surface interactions. Molecules are of theoretical interest because of their complex internal structure as opposed to atoms. For example the possibility of a chiral component of the CP interaction has been proposed when chiral molecules are located in front of chiral surfaces (for a review on chiral metamaterials see [220]). In addition, the effects of molecular orientation with respect to the planar surface (CP anisotropy) have been investigated theoretically [221], [222], [223]. Moreover, CP interactions with molecules are of importance in areas such as physical chemistry (studying molecular adsorption, desorption or catalysis) and could surprisingly be relevant in areas such as atmospheric physics [223], [224] and studies of the greenhouse effect. Up to now, experimental measurements of the CP interaction between molecules and surface have been scarce. After the first attempts to measure the mechanical displacement of a thermal molecular beams [63], [62], [225] in the 60' and 70's, and after the observation of surface-induced vibrational transitions induced in $N_2$ by the anisotropic part of the CP potential [68], some modern experiments have been performed aiming at measuring CP forces with an atomic force microscope (AFM) [226] or with the diffraction of macromolecule beams in nano-gratings [227]. More recently, molecular gases have been spectroscopically probed by high-resolution rovibrational selective reflection at micrometric distances away from dielectric surfaces using the newly developed quantum cascade laser (QCL) sources [228]. These experiments pave the way for state selective, precision spectroscopy of the CP interaction with molecules.

# 7. Acknowledgments

This work was supported by the Centre for Quantum Technologies funding Grant No. R-710-002-016-271, the Singapore Ministry of Education Academic Research Fund Tier3 Grant No. MOE2016-T3-1-006(S) and the Singapore Ministry of Education Academic Research Fund Tier 1 Grant No. RG160/19(S). The authors also acknowledge financial support from the ANR project SQUAT (Grant No. ANR-20-CE92-0006-0.1).

# 8. Conflict of interest

The authors have no conflicts to disclose.

# 9. Data availability

Data sharing is not applicable to this article as no new data were created or analyzed in this study.

## 10. References

[1] H. B. G. Casimir, "On the Attraction between Two Perfectly Conducting Plates," *Proc Kon Ned Akad Wet*, vol. 51, p. 793, 1948.

[2] H. B. G. Casimir and D. Polder, "The Influence of Retardation on the London-van der Waals Forces," *Phys. Rev.*, vol. 73, no. 4, pp. 360–372, Feb. 1948, doi: 10.1103/PhysRev.73.360.

[3] J. E. Lennard-Jones, "Processes of adsorption and diffusion on solid surfaces," *Trans Faraday Soc*, vol. 28, pp. 333–359, 1932.

[4] E. M. Lifshitz, "The theory of molecular attractive forces between solids," *Sov. Phys. JETP*, vol. 2, no. 1, p. 73, Jan. 1956.

[5] I. E. Dzyaloshinskii, E. M. Lifshitz, and L. P. Pitaevskii, "The general theory of van der Waals forces," *Adv. Phys.*, vol. 10, no. 38, pp. 165–209, Apr. 1961, doi: 10.1080/00018736100101281.

[6] A. D. McLachlan, "Retarded dispersion forces in dielectrics at finite temperatures," *Proc. R. Soc. Lond. Ser. Math. Phys. Sci.*, vol. 274, no. 1356, pp. 80–90, Jun. 1963, doi: 10.1098/rspa.1963.0115.

[7] J. M. Wylie and J. E. Sipe, "Quantum electrodynamics near an interface," *Phys. Rev. A*, vol. 30, no. 3, pp. 1185–1193, Sep. 1984, doi: 10.1103/PhysRevA.30.1185.

[8] J. M. Wylie and J. E. Sipe, "Quantum electrodynamics near an interface. II," *Phys. Rev. A*, vol. 32, no. 4, pp. 2030–2043, Oct. 1985, doi: 10.1103/PhysRevA.32.2030.

[9] V. Sandoghdar, C. I. Sukenik, E. A. Hinds, and S. Haroche, "Direct measurement of the van der Waals interaction between an atom and its images in a micron-sized cavity," *Phys. Rev. Lett.*, vol. 68, no. 23, pp. 3432–3435, Jun. 1992, doi: 10.1103/PhysRevLett.68.3432.

[10] C. I. Sukenik, M. G. Boshier, D. Cho, V. Sandoghdar, and E. A. Hinds, "Measurement of the Casimir-Polder force," *Phys. Rev. Lett.*, vol. 70, no. 5, pp. 560–563, Feb. 1993, doi: 10.1103/PhysRevLett.70.560.

[11] M. Oria, M. Chevrollier, D. Bloch, M. Fichet, and M. Ducloy, "Spectral Observation of Surface-Induced Van der Waals Attraction on Atomic Vapour," *Europhys. Lett. EPL*, vol. 14, no. 6, pp. 527–532, Mar. 1991, doi: 10.1209/0295-5075/14/6/005.

[12] D. Bloch and M. Ducloy, "Atom-wall interaction," in *Advances In Atomic, Molecular, and Optical Physics*, vol. 50, Elsevier, 2005, pp. 91–154. doi: 10.1016/S1049-250X(05)80008-4.

[13] A. Landragin, J.-Y. Courtois, G. Labeyrie, N. Vansteenkiste, C. I. Westbrook, and A. Aspect, "Measurement of the van der Waals Force in an Atomic Mirror," *Phys. Rev. Lett.*, vol. 77, no. 8, pp. 1464–1467, Aug. 1996, doi: 10.1103/PhysRevLett.77.1464.

[14] H. Bender, Ph. W. Courteille, C. Marzok, C. Zimmermann, and S. Slama, "Direct Measurement of Intermediate-Range Casimir-Polder Potentials," *Phys. Rev. Lett.*, vol. 104, no. 8, Feb. 2010, doi: 10.1103/PhysRevLett.104.083201.

[15] D. M. Harber, J. M. Obrecht, J. M. McGuirk, and E. A. Cornell, "Measurement of the Casimir-Polder force through center-of-mass oscillations of a Bose-Einstein condensate," *Phys. Rev. A*, vol. 72, no. 3, Sep. 2005, doi: 10.1103/PhysRevA.72.033610.

[16] P. Bushev *et al.*, "Forces between a Single Atom and Its Distant Mirror Image," *Phys. Rev. Lett.*, vol. 92, no. 22, Jun. 2004, doi: 10.1103/PhysRevLett.92.223602.

[17] A. Laliotis, T. P. de Silans, I. Maurin, M. Ducloy, and D. Bloch, "Casimir–Polder interactions in the presence of thermally excited surface modes," *Nat. Commun.*, vol. 5, Jul. 2014, doi: 10.1038/ncomms5364.

[18] A. V. Shchegrov, K. Joulain, R. Carminati, and J.-J. Greffet, "Near-Field Spectral Effects due to Electromagnetic Surface Excitations," *Phys. Rev. Lett.*, vol. 85, no. 7, pp. 1548–1551, Aug. 2000, doi: 10.1103/PhysRevLett.85.1548.

[19] J.-J. Greffet, R. Carminati, K. Joulain, J.-P. Mulet, S. Mainguy, and Y. Chen, "Coherent emission of light by thermal sources," *Nature*, vol. 416, no. 6876, pp. 61–64, Mar. 2002, doi: 10.1038/416061a.

[20] K. Joulain, J.-P. Mulet, F. Marquier, R. Carminati, and J.-J. Greffet, "Surface electromagnetic waves thermally excited: Radiative heat transfer, coherence properties and Casimir forces

revisited in the near field," *Surf. Sci. Rep.*, vol. 57, no. 3, pp. 59–112, May 2005, doi: 10.1016/j.surfrep.2004.12.002.

[21] D. E. Chang, J. S. Douglas, A. González-Tudela, C.-L. Hung, and H. J. Kimble, "*Colloquium* : Quantum matter built from nanoscopic lattices of atoms and photons," *Rev. Mod. Phys.*, vol. 90, no. 3, p. 031002, Aug. 2018, doi: 10.1103/RevModPhys.90.031002.

[22] A. M. Contreras Reyes and C. Eberlein, "Casimir-Polder interaction between an atom and a dielectric slab," *Phys. Rev. A*, vol. 80, no. 3, p. 032901, Sep. 2009, doi: 10.1103/PhysRevA.80.032901.

[23] W. Jhe and J. W. Kim, "Casimir-Polder energy shift of an atom near a metallic sphere," *Phys. Lett. A*, vol. 197, no. 3, pp. 192–196, Jan. 1995, doi: 10.1016/0375-9601(94)00966-S.

[24] J. L. Hemmerich *et al.*, "Impact of Casimir-Polder interaction on Poisson-spot diffraction at a dielectric sphere," *Phys. Rev. A*, vol. 94, no. 2, p. 023621, Aug. 2016, doi: 10.1103/PhysRevA.94.023621.

[25] V. V. Klimov, M. Ducloy, and V. S. Letokhov, "Radiative frequency shift and linewidth of an atom dipole in the vicinity of a dielectric microsphere," *J. Mod. Opt.*, vol. 43, no. 11, pp. 2251–2267, Nov. 1996, doi: 10.1080/09500349608232884.

[26] P. N. Prasad, *Nanophotonics*. John Wiley \& Sons, 2004.

[27] Á. M. Alhambra, A. Kempf, and E. Martín-Martínez, "Casimir forces on atoms in optical cavities," *Phys. Rev. A*, vol. 89, no. 3, p. 033835, Mar. 2014, doi: 10.1103/PhysRevA.89.033835.

[28] D. G. Baranov, M. Wersäll, J. Cuadra, T. J. Antosiewicz, and T. Shegai, "Novel Nanostructures and Materials for Strong Light–Matter Interactions," *ACS Photonics*, vol. 5, no. 1, pp. 24–42, Jan. 2018, doi: 10.1021/acsphotonics.7b00674.

[29] P. Lalanne, W. Yan, K. Vynck, C. Sauvan, and J.-P. Hugonin, "Light Interaction with Photonic and Plasmonic Resonances," *Laser Photonics Rev.*, vol. 12, no. 5, p. 1700113, May 2018, doi: 10.1002/lpor.201700113.

[30] J. M. Wylie and J. E. Sipe, "Quantum electrodynamics near an interface. II," *Phys. Rev. A*, vol. 32, no. 4, Art. no. 4, Oct. 1985, doi: 10.1103/PhysRevA.32.2030.

[31] S. Y. Buhmann, *Dispersion Forces I: Macroscopic Quantum Electrodynamics and Ground-State Casimir, Casimir-Polder and van der Waals Forces*. Springer-Verlag (Berlin), 2012.

[32] M.-P. Gorza and M. Ducloy, "Van der Waals interactions between atoms and dispersive surfaces at finite temperature," *Eur. Phys. J. D*, vol. 40, no. 3, Art. no. 3, Dec. 2006, doi: 10.1140/epjd/e2006-00239-3.

[33] L. Novotny and B. Hecht, *Principles of nano-optics*. Cambridge University Press, 2012.

[34] J. M. Wylie and J. E. Sipe, "Quantum electrodynamics near an interface," *Phys. Rev. A*, vol. 30, no. 3, Art. no. 3, Sep. 1984, doi: 10.1103/PhysRevA.30.1185.

[35] G. Barton, "Van der Waals shifts in an atom near absorptive dielectric mirrors," *Proc. R. Soc. Math. Phys. Eng. Sci.*, vol. 453, no. 1966, pp. 2461–2495, Nov. 1997, doi: 10.1098/rspa.1997.0132.

[36] S. Y. Buhmann and S. Scheel, "Thermal Casimir versus Casimir-Polder Forces: Equilibrium and Nonequilibrium Forces," *Phys. Rev. Lett.*, vol. 100, no. 25, Jun. 2008, doi: 10.1103/PhysRevLett.100.253201.

[37] M.-P. Gorza and M. Ducloy, "Van der Waals interactions between atoms and dispersive surfaces at finite temperature," *Eur. Phys. J. D*, vol. 40, no. 3, Art. no. 3, Dec. 2006, doi: 10.1140/epjd/e2006-00239-3.

[38] S. Y. Buhmann, L. Knöll, D.-G. Welsch, and H. T. Dung, "Casimir-Polder forces: A nonperturbative approach," *Phys. Rev. A*, vol. 70, no. 5, Art. no. 5, Nov. 2004, doi: 10.1103/PhysRevA.70.052117.

[39] M. Chevrollier, M. Fichet, M. Oria, G. Rahmat, D. Bloch, and M. Ducloy, "High resolution selective reflection spectroscopy as a probe of long-range surface interaction : measurement of the surface van der Waals attraction exerted on excited Cs atoms," *J. Phys. II*, vol. 2, no. 4, pp. 631–657, Apr. 1992, doi: 10.1051/jp2:1992108.

[40] O. S. Heavens, "Radiative Transition Probabilities of the Lower Excited States of the Alkali Metals," *J. Opt. Soc. Am.*, vol. 51, no. 10, p. 1058, Oct. 1961, doi: 10.1364/JOSA.51.001058.

[41] A. Lindgård and S. E. Nielsen, "Transition probabilities for the alkali isoelectronic sequences Li I, Na I, K I, Rb I, Cs I, Fr I," *At. Data Nucl. Data Tables*, vol. 19, no. 6, pp. 533–633, Jun. 1977, doi: 10.1016/0092-640X(77)90017-1.

[42] M. S. Safronova, U. I. Safronova, and C. W. Clark, "Magic wavelengths, matrix elements, polarizabilities, and lifetimes of Cs," *Phys. Rev. A*, vol. 94, no. 1, Jul. 2016, doi: 10.1103/PhysRevA.94.012505.

[43] E. A. Hinds and V. Sandoghdar, "Cavity QED level shifts of simple atoms," *Phys. Rev. A*, vol. 43, no. 1, pp. 398–403, Jan. 1991, doi: 10.1103/PhysRevA.43.398.

[44] J. M. Obrecht, R. J. Wild, M. Antezza, L. P. Pitaevskii, S. Stringari, and E. A. Cornell, "Measurement of the Temperature Dependence of the Casimir-Polder Force," *Phys. Rev. Lett.*, vol. 98, no. 6, Feb. 2007, doi: 10.1103/PhysRevLett.98.063201.

[45] A. Laliotis and M. Ducloy, "Casimir-Polder effect with thermally excited surfaces," *Phys. Rev. A*, vol. 91, no. 5, Art. no. 5, May 2015, doi: 10.1103/PhysRevA.91.052506.

[46] S. Y. Buhmann, *Dispersion Forces II: Many-Body Effects, Excited Atoms, Finite Temperature and Quantum Friction*. Springer-Verlag (Berlin), 2012.

[47] G. S. Agarwal, "Quantum electrodynamics in the presence of dielectrics and conductors. I. Electromagnetic-field response functions and black-body fluctuations in finite geometries," *Phys. Rev. A*, vol. 11, no. 1, pp. 230–242, Jan. 1975, doi: 10.1103/PhysRevA.11.230.

[48] S. Y. Buhmann, H. T. Dung, and D.-G. Welsch, "The van der Waals energy of atomic systems near absorbing and dispersing bodies," *J. Opt. B Quantum Semiclassical Opt.*, vol. 6, no. 3, Art. no. 3, Mar. 2004, doi: 10.1088/1464-4266/6/3/020.

[49] W. E. Lamb and R. C. Retherford, "Fine Structure of the Hydrogen Atom by a Microwave Method," *Phys. Rev.*, vol. 72, no. 3, pp. 241–243, Aug. 1947, doi: 10.1103/PhysRev.72.241.

[50] J. D. Jackson and R. F. Fox, "Classical Electrodynamics, 3rd ed," *Am. J. Phys.*, vol. 67, no. 9, pp. 841–842, Sep. 1999, doi: 10.1119/1.19136.

[51] M. Boustimi *et al.*, "Metastable rare gas atoms scattered by nano- and micro-slit transmission gratings," *Eur. Phys. J. D*, vol. 17, no. 2, pp. 141–144, Nov. 2001, doi: 10.1007/s100530170015.

[52] M. Boustimi, B. Viaris de Lesegno, J. Baudon, J. Robert, and M. Ducloy, "Atom Symmetry Break and Metastable Level Coupling in Rare Gas Atom-Surface van der Waals Interaction," *Phys. Rev. Lett.*, vol. 86, no. 13, pp. 2766–2769, Mar. 2001, doi: 10.1103/PhysRevLett.86.2766.

[53] S. Y. Buhmann and S. Scheel, "Thermal Casimir versus Casimir-Polder Forces: Equilibrium and Nonequilibrium Forces," *Phys. Rev. Lett.*, vol. 100, no. 25, Art. no. 25, Jun. 2008, doi: 10.1103/PhysRevLett.100.253201.

[54] A. Narayanaswamy and G. Chen, "Dyadic Green's functions and electromagnetic local density of states," *J. Quant. Spectrosc. Radiat. Transf.*, vol. 111, no. 12–13, pp. 1877–1884, Aug. 2010, doi: 10.1016/j.jqsrt.2009.12.008.

[55] C.-T. Tai, "Dyadic Green functions in electromagnetic theory," *IEEE*, 1994.

[56] H. Failache, S. Saltiel, A. Fischer, D. Bloch, and M. Ducloy, "Resonant Quenching of Gas-Phase Cs Atoms Induced by Surface Polaritons," *Phys. Rev. Lett.*, vol. 88, no. 24, Art. no. 24, May 2002, doi: 10.1103/PhysRevLett.88.243603.

[57] "[2101.00901] RETICOLO software for grating analysis." https://arxiv.org/abs/2101.00901 (accessed Jun. 16, 2021).

[58] C. Sauvan, J. P. Hugonin, I. S. Maksymov, and P. Lalanne, "Theory of the Spontaneous Optical Emission of Nanosize Photonic and Plasmon Resonators," *Phys. Rev. Lett.*, vol. 110, no. 23, p. 237401, Jun. 2013, doi: 10.1103/PhysRevLett.110.237401.

[59] E. Lassalle, N. Bonod, T. Durt, and B. Stout, "Interplay between spontaneous decay rates and Lamb shifts in open photonic systems," *Opt. Lett.*, vol. 43, no. 9, p. 1950, May 2018, doi: 10.1364/OL.43.001950.

[60] M. B. Doost, W. Langbein, and E. A. Muljarov, "Resonant-state expansion applied to three-dimensional open optical systems," *Phys. Rev. A*, vol. 90, no. 1, p. 013834, Jul. 2014, doi: 10.1103/PhysRevA.90.013834.

[61] E. A. Muljarov and W. Langbein, "Exact mode volume and Purcell factor of open optical systems," *Phys. Rev. B*, vol. 94, no. 23, p. 235438, Dec. 2016, doi: 10.1103/PhysRevB.94.235438.

[62] D. Raskin and P. Kusch, "Interaction between a Neutral Atomic or Molecular Beam and a Conducting Surface," *Phys. Rev.*, vol. 179, no. 3, pp. 712–721, Mar. 1969, doi: 10.1103/PhysRev.179.712.

[63] A. Shih, "van der Waals forces between a Cs atom or a CsCl molecule and metal or dielectric surfaces," *Phys. Rev. A*, vol. 9, no. 4, pp. 1507–1514, Apr. 1974, doi: 10.1103/PhysRevA.9.1507.

[64] A. Anderson, S. Haroche, E. A. Hinds, W. Jhe, and D. Meschede, "Measuring the van der Waals forces between a Rydberg atom and a metallic surface," *Phys. Rev. A*, vol. 37, no. 9, pp. 3594–3597, May 1988, doi: 10.1103/PhysRevA.37.3594.

[65] V. Sandoghdar, C. I. Sukenik, S. Haroche, and E. A. Hinds, "Spectroscopy of atoms confined to the single node of a standing wave in a parallel-plate cavity," *Phys. Rev. A*, vol. 53, no. 3, pp. 1919–1922, Mar. 1996, doi: 10.1103/PhysRevA.53.1919.

[66] T. Taillandier-Loize, J. Baudon, G. Dutier, F. Perales, M. Boustimi, and M. Ducloy, "Anisotropic atom-surface interactions in the Casimir-Polder regime," *Phys. Rev. A*, vol. 89, no. 5, p. 052514, May 2014, doi: 10.1103/PhysRevA.89.052514.

[67] M. Ducloy, "Quantum Optics of Atomic Systems Confined in a Dielectric Environment," in *Nanoscale Science and Technology*, N. García, M. Nieto-Vesperinas, and H. Rohrer, Eds. Dordrecht: Springer Netherlands, 1998, pp. 235–253. doi: 10.1007/978-94-011-5024-8_17.

[68] M. Boustimi *et al.*, "Surface-induced vibrational transition of metastable nitrogen molecules passing through a nano-slit grating," *Europhys. Lett. EPL*, vol. 56, no. 5, pp. 644–650, Dec. 2001, doi: 10.1209/epl/i2001-00569-0.

[69] J.-C. Karam *et al.*, "van der Waals - Zeeman transitions of metastable neon atoms passing through a micro-slit copper grating," *Europhys. Lett. EPL*, vol. 74, no. 1, pp. 36–42, Apr. 2006, doi: 10.1209/epl/i2005-10514-3.

[70] M. Hamamda *et al.*, "Atom-surface interaction at the nanometre scale: van der Waals-Zeeman transitions in a magnetic field," *EPL Europhys. Lett.*, vol. 98, no. 2, p. 23001, Apr. 2012, doi: 10.1209/0295-5075/98/23001.

[71] A. K. Mohapatra and C. S. Unnikrishnan, "Measurement of the van der Waals force using reflection of cold atoms from magnetic thin-film atom mirrors," *Europhys. Lett. EPL*, vol. 73, no. 6, pp. 839–845, Mar. 2006, doi: 10.1209/epl/i2005-10477-3.

[72] H. Bender *et al.*, "Probing Atom-Surface Interactions by Diffraction of Bose-Einstein Condensates," *Phys. Rev. X*, vol. 4, no. 1, Feb. 2014, doi: 10.1103/PhysRevX.4.011029.

[73] A. M. Contreras-Reyes, R. Guérout, P. A. M. Neto, D. A. R. Dalvit, A. Lambrecht, and S. Reynaud, "Casimir-Polder interaction between an atom and a dielectric grating," *Phys. Rev. A*, vol. 82, no. 5, p. 052517, Nov. 2010, doi: 10.1103/PhysRevA.82.052517.

[74] P. Schneeweiss *et al.*, "Dispersion forces between ultracold atoms and a carbon nanotube," *Nat. Nanotechnol.*, vol. 7, no. 8, pp. 515–519, Aug. 2012, doi: 10.1038/nnano.2012.93.

[75] E. Vetsch, D. Reitz, G. Sagué, R. Schmidt, S. T. Dawkins, and A. Rauschenbeutel, "Optical Interface Created by Laser-Cooled Atoms Trapped in the Evanescent Field Surrounding an Optical Nanofiber," *Phys. Rev. Lett.*, vol. 104, no. 20, May 2010, doi: 10.1103/PhysRevLett.104.203603.

[76] Y. Lin, I. Teper, C. Chin, and V. Vuletić, "Impact of the Casimir-Polder Potential and Johnson Noise on Bose-Einstein Condensate Stability Near Surfaces," *Phys. Rev. Lett.*, vol. 92, no. 5, Feb. 2004, doi: 10.1103/PhysRevLett.92.050404.

[77] D. J. Alton *et al.*, "Strong interactions of single atoms and photons near a dielectric boundary," *Nat. Phys.*, vol. 7, no. 2, pp. 159–165, Feb. 2011, doi: 10.1038/nphys1837.

[78] A. González-Tudela, C.-L. Hung, D. E. Chang, J. I. Cirac, and H. J. Kimble, "Subwavelength vacuum lattices and atom–atom interactions in two-dimensional photonic crystals," *Nat. Photonics*, vol. 9, no. 5, pp. 320–325, May 2015, doi: 10.1038/nphoton.2015.54.

[79] A. Goban *et al.*, "Atom–light interactions in photonic crystals," *Nat. Commun.*, vol. 5, p. 3808, May 2014.

[80] C.-L. Hung, S. M. Meenehan, D. E. Chang, O. Painter, and H. J. Kimble, "Trapped atoms in one-dimensional photonic crystals," *New J. Phys.*, vol. 15, no. 8, p. 083026, Aug. 2013, doi: 10.1088/1367-2630/15/8/083026.

[81] M. Antezza, L. P. Pitaevskii, and S. Stringari, "Effect of the Casimir-Polder force on the collective oscillations of a trapped Bose-Einstein condensate," *Phys. Rev. A*, vol. 70, no. 5, p. 053619, Nov. 2004, doi: 10.1103/PhysRevA.70.053619.

[82] J. M. McGuirk, D. M. Harber, J. M. Obrecht, and E. A. Cornell, "Alkali-metal adsorbate polarization on conducting and insulating surfaces probed with Bose-Einstein condensates," *Phys. Rev. A*, vol. 69, no. 6, p. 062905, Jun. 2004, doi: 10.1103/PhysRevA.69.062905.

[83] J. M. Obrecht, R. J. Wild, and E. A. Cornell, "Measuring electric fields from surface contaminants with neutral atoms," *Phys. Rev. A*, vol. 75, no. 6, Jun. 2007, doi: 10.1103/PhysRevA.75.062903.

[84] M. Antezza, L. P. Pitaevskii, and S. Stringari, "New Asymptotic Behavior of the Surface-Atom Force out of Thermal Equilibrium," *Phys. Rev. Lett.*, vol. 95, no. 11, Sep. 2005, doi: 10.1103/PhysRevLett.95.113202.

[85] M. A. Wilson *et al.*, "Vacuum-Field Level Shifts in a Single Trapped Ion Mediated by a Single Distant Mirror," *Phys. Rev. Lett.*, vol. 91, no. 21, p. 213602, Nov. 2003, doi: 10.1103/PhysRevLett.91.213602.

[86] F. Shimizu, "Specular Reflection of Very Slow Metastable Neon Atoms from a Solid Surface," *Phys. Rev. Lett.*, vol. 86, no. 6, pp. 987–990, Feb. 2001, doi: 10.1103/PhysRevLett.86.987.

[87] T. A. Pasquini *et al.*, "Quantum Reflection from a Solid Surface at Normal Incidence," *Phys. Rev. Lett.*, vol. 93, no. 22, p. 223201, Nov. 2004, doi: 10.1103/PhysRevLett.93.223201.

[88] B. S. Zhao, G. Meijer, and W. Schollkopf, "Quantum Reflection of He2 Several Nanometers Above a Grating Surface," *Science*, vol. 331, no. 6019, pp. 892–894, Feb. 2011, doi: 10.1126/science.1200911.

[89] P.-P. Crépin, G. Dufour, R. Guérout, A. Lambrecht, and S. Reynaud, "Casimir-Polder shifts on quantum levitation states," *Phys. Rev. A*, vol. 95, no. 3, p. 032501, Mar. 2017, doi: 10.1103/PhysRevA.95.032501.

[90] M. Silvestre, T. Cysne, D. Szilard, F. A. Pinheiro, and C. Farina, "Tuning quantum reflection in graphene with an external magnetic field," *Phys. Rev. A*, vol. 100, p. 033605, Sep. 2019.

[91] P. P. Abrantes, T. P. Cysne, D. Szilard, F. S. S. Rosa, F. A. Pinheiro, and C. Farina, "Probing topological phase transitions via quantum reflection in the graphene family materials," *Phys. Rev. B*, vol. 104, p. 075409, Aug. 2021.

[92] F. Sorrentino *et al.*, "Quantum sensor for atom-surface interactions below 10 μm," *Phys. Rev. A*, vol. 79, no. 1, p. 013409, Jan. 2009, doi: 10.1103/PhysRevA.79.013409.

[93] A. Derevianko, B. Obreshkov, and V. A. Dzuba, "Mapping Out Atom-Wall Interaction with Atomic Clocks," *Phys. Rev. Lett.*, vol. 103, no. 13, p. 133201, Sep. 2009, doi: 10.1103/PhysRevLett.103.133201.

[94] S. Pelisson, R. Messina, M.-C. Angonin, and P. Wolf, "Dynamical aspects of atom interferometry in an optical lattice in proximity to a surface," *Phys. Rev. A*, vol. 86, no. 1, p. 013614, Jul. 2012, doi: 10.1103/PhysRevA.86.013614.

[95] A. Maury, M. Donaire, M.-P. Gorza, A. Lambrecht, and R. Guérout, "Surface-modified Wannier-Stark states in a one-dimensional optical lattice," *Phys. Rev. A*, vol. 94, no. 5, p. 053602, Nov. 2016, doi: 10.1103/PhysRevA.94.053602.

[96] R. E. Grisenti, W. Schöllkopf, J. P. Toennies, G. C. Hegerfeldt, and T. Köhler, "Determination of Atom-Surface van der Waals Potentials from Transmission-Grating Diffraction Intensities," *Phys. Rev. Lett.*, vol. 83, no. 9, pp. 1755–1758, Aug. 1999, doi: 10.1103/PhysRevLett.83.1755.

[97] R. Brühl *et al.*, "The van der Waals potential between metastable atoms and solid surfaces: Novel diffraction experiments *vs.* theory," *Europhys. Lett. EPL*, vol. 59, no. 3, pp. 357–363, Aug. 2002, doi: 10.1209/epl/i2002-00202-4.

[98] J.-C. Karam *et al.*, "Atom diffraction with a 'natural' metastable atom nozzle beam," *J. Phys. B At. Mol. Opt. Phys.*, vol. 38, no. 15, pp. 2691–2700, Aug. 2005, doi: 10.1088/0953-4075/38/15/009.

[99] J. D. Perreault, A. D. Cronin, and T. A. Savas, "Using atomic diffraction of Na from material gratings to measure atom-surface interactions," *Phys. Rev. A*, vol. 71, no. 5, p. 053612, May 2005, doi: 10.1103/PhysRevA.71.053612.

[100] T. Taillandier-Loize et al., "A simple velocity-tunable pulsed atomic source of slow metastable argon," *J. Phys. Appl. Phys.*, vol. 49, no. 13, p. 135503, Apr. 2016, doi: 10.1088/0022-3727/49/13/135503.

[101] C. Garcion et al., "Intermediate-range Casimir-Polder interaction probed by high-order slow atom diffraction," Apr. 2021. Accessed: May 14, 2021. [Online]. Available: https://hal.archives-ouvertes.fr/hal-03200194

[102] J. D. Perreault and A. D. Cronin, "Observation of Atom Wave Phase Shifts Induced by Van Der Waals Atom-Surface Interactions," *Phys. Rev. Lett.*, vol. 95, no. 13, p. 133201, Sep. 2005, doi: 10.1103/PhysRevLett.95.133201.

[103] S. Lepoutre et al., "Dispersive atom interferometry phase shifts due to atom-surface interactions," *EPL Europhys. Lett.*, vol. 88, no. 2, p. 20002, Oct. 2009, doi: 10.1209/0295-5075/88/20002.

[104] S. Lepoutre et al., "Atom interferometry measurement of the atom-surface van der Waals interaction," *Eur. Phys. J. D*, vol. 62, no. 3, pp. 309–325, May 2011, doi: 10.1140/epjd/e2011-10584-7.

[105] J.-L. Cojan, "Contribution à l'étude de la réflexion sélective sur les vapeurs de mercure de la radiation de résonance du mercure," *Ann. Phys.*, vol. 12, no. 9, pp. 385–440, 1954, doi: 10.1051/anphys/195412090385.

[106] J. P. Woerdman and M. F. H. Schuurmans, "Spectral narrowing of selective reflection from sodium vapour," *Opt. Commun.*, vol. 14, no. 2, pp. 248–251, Jun. 1975, doi: 10.1016/0030-4018(75)90226-6.

[107] A. M. Akulshin et al., "Collisional Broadening of Intra-Doppler Resonances of Selective Reflection on the D2 Line of Cesium," *Jetp Letters*, pp. 303–307, 1975.

[108] M. Ducloy and M. Fichet, "General theory of frequency modulated selective reflection. Influence of atom surface interactions," *J. Phys. II*, vol. 1, no. 12, pp. 1429–1446, Dec. 1991, doi: 10.1051/jp2:1991160.

[109] M. Chevrollier, D. Bloch, G. Rahmat, and M. Ducloy, "Van der Waals-induced spectral distortions in selective-reflection spectroscopy of Cs vapor: the strong atom–surface interaction regime," *Opt. Lett.*, vol. 16, no. 23, p. 1879, Dec. 1991, doi: 10.1364/OL.16.001879.

[110] A. M. Akulshin, A. A. Celikov, V. A. Sautenkov, T. A. Vartanian, and V. L. Velichansky, "Intensity and concentration dependence of Doppler-free resonance in selective reflection," *Opt. Commun.*, vol. 85, no. 1, pp. 21–25, Aug. 1991, doi: 10.1016/0030-4018(91)90045-F.

[111] V. Vuletić, V. A. Sautenkov, C. Zimmermann, and T. W. Hänsch, "Measurement of cesium resonance line self-broadening and shift with doppler-free selective reflection spectroscopy," *Opt. Commun.*, vol. 99, no. 3–4, pp. 185–190, Jun. 1993, doi: 10.1016/0030-4018(93)90076-H.

[112] J. Guo, J. Cooper, and A. Gallagher, "Selective reflection from a dense atomic vapor," *Phys. Rev. A*, vol. 53, no. 2, Art. no. 2, Feb. 1996, doi: 10.1103/PhysRevA.53.1130.

[113] P. Wang, A. Gallagher, and J. Cooper, "Selective reflection by Rb," *Phys. Rev. A*, vol. 56, no. 2, pp. 1598–1606, Aug. 1997, doi: 10.1103/PhysRevA.56.1598.

[114] H. Failache, S. Saltiel, M. Fichet, D. Bloch, and M. Ducloy, "Resonant coupling in the van der Waals interaction between an excited alkali atom and a dielectric surface: an experimental study via stepwise selective reflection spectroscopy," *Eur. Phys. J. - At. Mol. Opt. Phys.*, vol. 23, no. 2, pp. 237–255, May 2003, doi: 10.1140/epjd/e2003-00098-4.

[115] H. Failache, S. Saltiel, M. Fichet, D. Bloch, and M. Ducloy, "Resonant van der Waals Repulsion between Excited Cs Atoms and Sapphire Surface," *Phys. Rev. Lett.*, vol. 83, no. 26, pp. 5467–5470, Dec. 1999, doi: 10.1103/PhysRevLett.83.5467.

[116] S. Briaudeau, D. Bloch, and M. Ducloy, "Detection of slow atoms in laser spectroscopy of a thin vapor film," *Europhys. Lett. EPL*, vol. 35, no. 5, pp. 337–342, Aug. 1996, doi: 10.1209/epl/i1996-00116-1.

[117] S. Briaudeau, S. Saltiel, G. Nienhuis, D. Bloch, and M. Ducloy, "Coherent Doppler narrowing in a thin vapor cell: Observation of the Dicke regime in the optical domain," *Phys. Rev. A*, vol. 57, no. 5, pp. R3169–R3172, May 1998, doi: 10.1103/PhysRevA.57.R3169.

[118] G. Dutier *et al.*, "Collapse and revival of a Dicke-type coherent narrowing in a sub-micron thick vapor cell transmission spectroscopy," *Europhys. Lett. EPL*, vol. 63, no. 1, pp. 35–41, Jul. 2003, doi: 10.1209/epl/i2003-00474-0.

[119] D. Sarkisyan, D. Bloch, A. Papoyan, and M. Ducloy, "Sub-Doppler spectroscopy by sub-micron thin Cs vapour layer," *Opt. Commun.*, vol. 200, no. 1–6, pp. 201–208, Dec. 2001, doi: 10.1016/S0030-4018(01)01604-2.

[120] M. Fichet *et al.*, "Exploring the van der Waals atom-surface attraction in the nanometric range," *Europhys. Lett. EPL*, vol. 77, no. 5, p. 54001, Mar. 2007, doi: 10.1209/0295-5075/77/54001.

[121] A. Laliotis *et al.*, "Testing the distance-dependence of the van der Waals interaction between an atom and a surface through spectroscopy in a vapor nanocell," Mar. 2007, pp. 660406-660406–11. doi: 10.1117/12.726798.

[122] J. C. de A. Carvalho, P. Pedri, M. Ducloy, and A. Laliotis, "Retardation effects in spectroscopic measurements of the Casimir-Polder interaction," *Phys. Rev. A*, vol. 97, no. 2, Feb. 2018, doi: 10.1103/PhysRevA.97.023806.

[123] G. Dutier, S. Saltiel, D. Bloch, and M. Ducloy, "Revisiting optical spectroscopy in a thin vapor cell: mixing of reflection and transmission as a Fabry–Perot microcavity effect," *J. Opt. Soc. Am. B*, vol. 20, no. 5, p. 793, May 2003, doi: 10.1364/JOSAB.20.000793.

[124] K. A. Whittaker, J. Keaveney, I. G. Hughes, A. Sargsyan, D. Sarkisyan, and C. S. Adams, "Optical Response of Gas-Phase Atoms at Less than λ / 80 from a Dielectric Surface," *Phys. Rev. Lett.*, vol. 112, no. 25, Jun. 2014, doi: 10.1103/PhysRevLett.112.253201.

[125] K. A. Whittaker, J. Keaveney, I. G. Hughes, A. Sargsyan, D. Sarkisyan, and C. S. Adams, "Spectroscopic detection of atom-surface interactions in an atomic-vapor layer with nanoscale thickness," *Phys. Rev. A*, vol. 92, no. 5, Nov. 2015, doi: 10.1103/PhysRevA.92.052706.

[126] D. Bloch, "Comment on 'Optical Response of Gas-Phase Atoms at Less than λ / 80 from a Dielectric Surface,'" *Phys. Rev. Lett.*, vol. 114, no. 4, p. 049301, Jan. 2015, doi: 10.1103/PhysRevLett.114.049301.

[127] T. Peyrot *et al.*, "Measurement of the atom-surface van der Waals interaction by transmission spectroscopy in a wedged nanocell," *Phys. Rev. A*, vol. 100, no. 2, p. 022503, Aug. 2019, doi: 10.1103/PhysRevA.100.022503.

[128] A. Laliotis *et al.*, "Selective reflection spectroscopy at the interface between a calcium fluoride window and Cs vapour," *Appl. Phys. B*, vol. 90, no. 3–4, pp. 415–420, Mar. 2008, doi: 10.1007/s00340-007-2927-9.

[129] M. Fichet, F. Schuller, D. Bloch, and M. Ducloy, "van der Waals interactions between excited-state atoms and dispersive dielectric surfaces," *Phys. Rev. A*, vol. 51, no. 2, pp. 1553–1564, Feb. 1995, doi: 10.1103/PhysRevA.51.1553.

[130] Y. Liu and X. Zhang, "Metamaterials: a new frontier of science and technology," *Chem. Soc. Rev.*, vol. 40, no. 5, p. 2494, 2011, doi: 10.1039/c0cs00184h.

[131] S. A. Aljunid, E. A. Chan, G. Adamo, M. Ducloy, D. Wilkowski, and N. I. Zheludev, "Atomic Response in the Near-Field of Nanostructured Plasmonic Metamaterial," *Nano Lett.*, vol. 16, no. 5, pp. 3137–3141, May 2016, doi: 10.1021/acs.nanolett.6b00446.

[132] E. A. Chan, S. A. Aljunid, G. Adamo, A. Laliotis, M. Ducloy, and D. Wilkowski, "Tailoring optical metamaterials to tune the atom-surface Casimir-Polder interaction," *Sci. Adv.*, vol. 4, no. 2, Art. no. 2, Feb. 2018, doi: 10.1126/sciadv.aao4223.

[133] E. A. Chan, S. A. Aljunid, G. Adamo, N. I. Zheludev, M. Ducloy, and D. Wilkowski, "Coupling of atomic quadrupole transitions with resonant surface plasmons," *Phys. Rev. A*, vol. 99, no. 6, p. 063801, Jun. 2019, doi: 10.1103/PhysRevA.99.063801.

[134] E. A. Chan, G. Adamo, S. A. Aljunid, M. Ducloy, N. Zheludev, and D. Wilkowski, "Plasmono-atomic interactions on a fiber tip," *Appl. Phys. Lett.*, vol. 116, no. 18, p. 183101, May 2020, doi: 10.1063/1.5142411.

[135] D. E. Chang, K. Sinha, J. M. Taylor, and H. J. Kimble, "Trapping atoms using nanoscale quantum vacuum forces," *Nat. Commun.*, vol. 5, Jul. 2014, doi: 10.1038/ncomms5343.

[136] T. Passerat de Silans *et al.*, "Experimental observations of temperature effects in the near-field regime of the Casimir–Polder interaction," *Laser Phys.*, vol. 24, no. 7, p. 074009, Jul. 2014, doi: 10.1088/1054-660X/24/7/074009.

[137] J. C. de Aquino Carvalho, "Interaction Casimir-Polder entre atome de césium et surface de saphir thermiquement émissive." UNIVERSITE PARIS 13, Jul. 09, 2018.

[138] E. A. Hinds, K. S. Lai, and M. Schnell, "Atoms in micron-sized metallic and dielectric waveguides," *Philos. Trans. R. Soc. Math. Phys. Eng. Sci.*, vol. 355, no. 1733, pp. 2353–2365, Dec. 1997, doi: 10.1098/rsta.1997.0132.

[139] K. S. Lai and E. A. Hinds, "Blackbody Excitation of an Atom Controlled by a Tunable Cavity," *Phys. Rev. Lett.*, vol. 81, no. 13, pp. 2671–2674, Sep. 1998, doi: 10.1103/PhysRevLett.81.2671.

[140] S. Å. Ellingsen, S. Y. Buhmann, and S. Scheel, "Temperature-Independent Casimir-Polder Forces Despite Large Thermal Photon Numbers," *Phys. Rev. Lett.*, vol. 104, no. 22, Jun. 2010, doi: 10.1103/PhysRevLett.104.223003.

[141] J. A. Crosse, S. Å. Ellingsen, K. Clements, S. Y. Buhmann, and S. Scheel, "Thermal Casimir-Polder shifts in Rydberg atoms near metallic surfaces," *Phys. Rev. A*, vol. 82, no. 1, Jul. 2010, doi: 10.1103/PhysRevA.82.010901.

[142] A. S. Barker, "Infrared Lattice Vibrations and Dielectric Dispersion in Corundum," *Phys. Rev.*, vol. 132, no. 4, pp. 1474–1481, Nov. 1963, doi: 10.1103/PhysRev.132.1474.

[143] T. Passerat de Silans *et al.*, "Temperature dependence of the dielectric permittivity of $CaF_2$, $BaF_2$ and $Al_2O_3$: application to the prediction of a temperature-dependent van der Waals surface interaction exerted onto a neighbouring $Cs(8P_{3/2})$ atom," *J. Phys. Condens. Matter*, vol. 21, no. 25, p. 255902, Jun. 2009, doi: 10.1088/0953-8984/21/25/255902.

[144] W. G. Spitzer and D. A. Kleinman, "Infrared Lattice Bands of Quartz," *Phys. Rev.*, vol. 121, no. 5, pp. 1324–1335, Mar. 1961, doi: 10.1103/PhysRev.121.1324.

[145] S. K. Andersson, M. E. Thomas, and C. E. Hoffman, "Multiphonon contribution to the reststrahlen band of $BaF2$," *Infrared Phys. Technol.*, vol. 39, no. 1, pp. 47–54, Feb. 1998, doi: 10.1016/S1350-4495(97)00044-3.

[146] P. Wolf, P. Lemonde, A. Lambrecht, S. Bize, A. Landragin, and A. Clairon, "From optical lattice clocks to the measurement of forces in the Casimir regime," *Phys. Rev. A*, vol. 75, no. 6, Art. no. 6, Jun. 2007, doi: 10.1103/PhysRevA.75.063608.

[147] S. Saltiel, D. Bloch, and M. Ducloy, "A tabulation and critical analysis of the wavelength-dependent dielectric image coefficient for the interaction exerted by a surface onto a neighbouring excited atom," *Opt. Commun.*, vol. 265, no. 1, pp. 220–233, Sep. 2006, doi: 10.1016/j.optcom.2006.03.034.

[148] Joseph T. Lamberti and N. T. Saunders, "COMPATIBILITY OF CESIUM VAPOR WITH SELECTED MATERIALS AT TEMPERATURES TO 1200° F." NASA technical note, Aug. 1963.

[149] J. C. de Aquino Carvalho, I. Maurin, H. Failache, D. Bloch, and A. Laliotis, "Velocity preserving transfer between highly excited atomic states: black body radiation and collisions," *J. Phys. B At. Mol. Opt. Phys.*, vol. 54, no. 3, p. 035203, Feb. 2021, doi: 10.1088/1361-6455/abd532.

[150] S. Tojo, M. Hasuo, and T. Fujimoto, "Absorption Enhancement of an Electric Quadrupole Transition of Cesium Atoms in an Evanescent Field," *Phys. Rev. Lett.*, vol. 92, no. 5, Art. no. 5, Feb. 2004, doi: 10.1103/PhysRevLett.92.053001.

[151] V. V. Klimov and M. Ducloy, "Allowed and forbidden transitions in an atom placed near an ideally conducting cylinder," *Phys. Rev. A*, vol. 62, no. 4, p. 043818, Sep. 2000, doi: 10.1103/PhysRevA.62.043818.

[152] A. M. Kern and O. J. F. Martin, "Strong enhancement of forbidden atomic transitions using plasmonic nanostructures," *Phys. Rev. A*, vol. 85, no. 2, p. 022501, Feb. 2012, doi: 10.1103/PhysRevA.85.022501.

[153] K. Shibata, S. Tojo, and D. Bloch, "Excitation enhancement in electric multipole transitions near a nanoedge," *Opt. Express*, vol. 25, no. 8, Art. no. 8, Apr. 2017, doi: 10.1364/OE.25.009476.

[154] N. Rivera, I. Kaminer, B. Zhen, J. D. Joannopoulos, and M. Soljačić, "Shrinking light to allow forbidden transitions on the atomic scale," *Science*, vol. 353, no. 6296, pp. 263–269, Jul. 2016, doi: 10.1126/science.aaf6308.

[155] C. Huo, Z. Yan, X. Song, and H. Zeng, "2D materials via liquid exfoliation: a review on fabrication and applications," *Sci. Bull.*, vol. 60, no. 23, pp. 1994–2008, Dec. 2015, doi: 10.1007/s11434-015-0936-3.

[156] A. M. Urbas *et al.*, "Roadmap on optical metamaterials," *J. Opt.*, vol. 18, no. 9, p. 093005, Sep. 2016, doi: 10.1088/2040-8978/18/9/093005.

[157] M. I. Stockman *et al.*, "Roadmap on plasmonics," *J. Opt.*, vol. 20, no. 4, p. 043001, Apr. 2018, doi: 10.1088/2040-8986/aaa114.

[158] A. Poddubny, I. Iorsh, P. Belov, and Y. Kivshar, "Hyperbolic metamaterials," *Nat. Photonics*, vol. 7, no. 12, pp. 948–957, Dec. 2013, doi: 10.1038/nphoton.2013.243.

[159] O. Takayama and A. V. Lavrinenko, "Optics with hyperbolic materials [Invited]," *J. Opt. Soc. Am. B*, vol. 36, no. 8, p. F38, Aug. 2019, doi: 10.1364/JOSAB.36.000F38.

[160] L. Stern, M. Grajower, and U. Levy, "Fano resonances and all-optical switching in a resonantly coupled plasmonic–atomic system," *Nat. Commun.*, vol. 5, no. 1, p. 4865, Dec. 2014, doi: 10.1038/ncomms5865.

[161] C. Stehle, C. Zimmermann, and S. Slama, "Cooperative coupling of ultracold atoms and surface plasmons," *Nat. Phys.*, vol. 10, no. 12, pp. 937–942, Dec. 2014, doi: 10.1038/nphys3129.

[162] V. G. Kravets, A. V. Kabashin, W. L. Barnes, and A. N. Grigorenko, "Plasmonic Surface Lattice Resonances: A Review of Properties and Applications," *Chem. Rev.*, vol. 118, no. 12, pp. 5912–5951, Jun. 2018, doi: 10.1021/acs.chemrev.8b00243.

[163] D. S. Dovzhenko, S. V. Ryabchuk, Yu. P. Rakovich, and I. R. Nabiev, "Light–matter interaction in the strong coupling regime: configurations, conditions, and applications," *Nanoscale*, vol. 10, no. 8, pp. 3589–3605, 2018, doi: 10.1039/C7NR06917K.

[164] R. Chikkaraddy *et al.*, "Single-molecule strong coupling at room temperature in plasmonic nanocavities," *Nature*, vol. 535, no. 7610, pp. 127–130, Jul. 2016, doi: 10.1038/nature17974.

[165] H. Groß, J. M. Hamm, T. Tufarelli, O. Hess, and B. Hecht, "Near-field strong coupling of single quantum dots," *Sci. Adv.*, vol. 4, no. 3, p. eaar4906, Mar. 2018, doi: 10.1126/sciadv.aar4906.

[166] H. Safari, S. Y. Buhmann, D.-G. Welsch, and H. T. Dung, "Body-assisted van der Waals interaction between two atoms," *Phys. Rev. A*, vol. 74, no. 4, p. 042101, Oct. 2006, doi: 10.1103/PhysRevA.74.042101.

[167] H. Safari and M. R. Karimpour, "Body-Assisted van der Waals Interaction between Excited Atoms," *Phys. Rev. Lett.*, vol. 114, no. 1, p. 013201, Jan. 2015, doi: 10.1103/PhysRevLett.114.013201.

[168] Q.-Z. Yuan, C.-H. Yuan, and W. Zhang, "Near-surface effect on interatomic resonance interaction," *Phys. Rev. A*, vol. 93, no. 3, p. 032517, Mar. 2016, doi: 10.1103/PhysRevA.93.032517.

[169] W. D. Newman *et al.*, "Observation of long-range dipole-dipole interactions in hyperbolic metamaterials," *Sci. Adv.*, vol. 4, no. 10, p. eaar5278, Oct. 2018, doi: 10.1126/sciadv.aar5278.

[170] L. M. Woods, D. A. R. Dalvit, A. Tkatchenko, P. Rodriguez-Lopez, A. W. Rodriguez, and R. Podgornik, "Materials perspective on Casimir and van der Waals interactions," *Rev. Mod. Phys.*, vol. 88, no. 4, p. 045003, Nov. 2016, doi: 10.1103/RevModPhys.88.045003.

[171] A. G. Grushin and A. Cortijo, "Tunable Casimir Repulsion with Three-Dimensional Topological Insulators," *Phys. Rev. Lett.*, vol. 106, no. 2, p. 020403, Jan. 2011, doi: 10.1103/PhysRevLett.106.020403.

[172] P. Rodriguez-Lopez and A. G. Grushin, "Repulsive Casimir Effect with Chern Insulators," *Phys. Rev. Lett.*, vol. 112, no. 5, p. 056804, Feb. 2014, doi: 10.1103/PhysRevLett.112.056804.

[173] K. A. Milton, E. K. Abalo, P. Parashar, N. Pourtolami, I. Brevik, and S. Å. Ellingsen, "Repulsive Casimir and Casimir–Polder forces," *J. Phys. Math. Theor.*, vol. 45, no. 37, p. 374006, Sep. 2012, doi: 10.1088/1751-8113/45/37/374006.

[174] K. V. Shajesh and M. Schaden, "Repulsive long-range forces between anisotropic atoms and dielectrics," *Phys. Rev. A*, vol. 85, no. 1, p. 012523, Jan. 2012, doi: 10.1103/PhysRevA.85.012523.

[175] P. P. Abrantes, Y. França, F. S. S. da Rosa, C. Farina, and R. de Melo e Souza, "Repulsive van der Waals interaction between a quantum particle and a conducting toroid," *Phys. Rev. A*, vol. 98, no. 1, p. 012511, Jul. 2018, doi: 10.1103/PhysRevA.98.012511.

[176] S. Y. Buhmann, V. N. Marachevsky, and S. Scheel, "Impact of anisotropy on the interaction of an atom with a one-dimensional nano-grating," *Int. J. Mod. Phys. A*, vol. 31, no. 02n03, p. 1641029, Jan. 2016, doi: 10.1142/S0217751X16410293.

[177] M. Levin, A. P. McCauley, A. W. Rodriguez, M. T. H. Reid, and S. G. Johnson, "Casimir Repulsion between Metallic Objects in Vacuum," *Phys. Rev. Lett.*, vol. 105, no. 9, p. 090403, Aug. 2010, doi: 10.1103/PhysRevLett.105.090403.

[178] C. Eberlein and R. Zietal, "Casimir-Polder interaction between a polarizable particle and a plate with a hole," *Phys. Rev. A*, vol. 83, no. 5, p. 052514, May 2011, doi: 10.1103/PhysRevA.83.052514.

[179] J. J. Marchetta, P. Parashar, and K. V. Shajesh, "Geometrical dependence in Casimir-Polder repulsion," *ArXiv201111870 Quant-Ph*, Nov. 2020, Accessed: Mar. 15, 2021. [Online]. Available: http://arxiv.org/abs/2011.11870

[180] K. A. Milton, E. K. Abalo, P. Parashar, N. Pourtolami, I. Brevik, and S. Å. Ellingsen, "Casimir-Polder repulsion near edges: Wedge apex and a screen with an aperture," *Phys. Rev. A*, vol. 83, no. 6, p. 062507, Jun. 2011, doi: 10.1103/PhysRevA.83.062507.

[181] P. S. Venkataram, S. Molesky, P. Chao, and A. W. Rodriguez, "Fundamental limits to attractive and repulsive Casimir-Polder forces," *Phys. Rev. A*, vol. 101, no. 5, p. 052115, May 2020, doi: 10.1103/PhysRevA.101.052115.

[182] Q.-Z. Yuan, "Repulsive Casimir-Polder potential by a negative reflecting surface," *Phys. Rev. A*, vol. 92, no. 1, p. 012522, Jul. 2015, doi: 10.1103/PhysRevA.92.012522.

[183] A. Sambale, D.-G. Welsch, H. T. Dung, and S. Y. Buhmann, "van der Waals interaction and spontaneous decay of an excited atom in a superlens-type geometry," *Phys. Rev. A*, vol. 78, no. 5, p. 053828, Nov. 2008, doi: 10.1103/PhysRevA.78.053828.

[184] T. H. Boyer, "Recalculations of Long-Range van der Waals Potentials," *Phys. Rev.*, vol. 180, no. 1, pp. 19–24, Apr. 1969, doi: 10.1103/PhysRev.180.19.

[185] B.-S. Skagerstam, P. K. Rekdal, and A. H. Vaskinn, "Theory of Casimir-Polder forces," *Phys. Rev. A*, vol. 80, no. 2, p. 022902, Aug. 2009, doi: 10.1103/PhysRevA.80.022902.

[186] G. Bimonte, G. L. Klimchitskaya, and V. M. Mostepanenko, "Impact of magnetic properties on atom-wall interactions," *Phys. Rev. A*, vol. 79, no. 4, p. 042906, Apr. 2009, doi: 10.1103/PhysRevA.79.042906.

[187] K. Sinha, "Repulsive vacuum-induced forces on a magnetic particle," *Phys. Rev. A*, vol. 97, no. 3, p. 032513, Mar. 2018, doi: 10.1103/PhysRevA.97.032513.

[188] S. Y. Buhmann, D.-G. Welsch, and T. Kampf, "Ground-state van der Waals forces in planar multilayer magnetodielectrics," *Phys. Rev. A*, vol. 72, no. 3, p. 032112, Sep. 2005, doi: 10.1103/PhysRevA.72.032112.

[189] D. R. Smith, W. J. Padilla, D. C. Vier, S. C. Nemat-Nasser, and S. Schultz, "Composite Medium with Simultaneously Negative Permeability and Permittivity," *Phys. Rev. Lett.*, vol. 84, no. 18, pp. 4184–4187, May 2000, doi: 10.1103/PhysRevLett.84.4184.

[190] W. J. Padilla, A. J. Taylor, C. Highstrete, M. Lee, and R. D. Averitt, "Dynamical Electric and Magnetic Metamaterial Response at Terahertz Frequencies," *Phys. Rev. Lett.*, vol. 96, no. 10, p. 107401, Mar. 2006, doi: 10.1103/PhysRevLett.96.107401.

[191] S. Zhang, W. Fan, N. C. Panoiu, K. J. Malloy, R. M. Osgood, and S. R. J. Brueck, "Experimental Demonstration of Near-Infrared Negative-Index Metamaterials," *Phys. Rev. Lett.*, vol. 95, no. 13, p. 137404, Sep. 2005, doi: 10.1103/PhysRevLett.95.137404.

[192] A. A. Banishev, C.-C. Chang, G. L. Klimchitskaya, V. M. Mostepanenko, and U. Mohideen, "Measurement of the gradient of the Casimir force between a nonmagnetic gold sphere and a

magnetic nickel plate," *Phys. Rev. B*, vol. 85, no. 19, p. 195422, May 2012, doi: 10.1103/PhysRevB.85.195422.

[193] A. Sambale, S. Y. Buhmann, H. T. Dung, and D.-G. Welsch, "Resonant Casimir–Polder forces in planar meta-materials," *Phys. Scr.*, vol. T135, p. 014019, Jul. 2009, doi: 10.1088/0031-8949/2009/T135/014019.

[194] D. Vanderbilt, *Berry Phases in Electronic Structure Theory: Electric Polarization, Orbital Magnetization and Topological Insulators*, 1st ed. Cambridge University Press, 2018. doi: 10.1017/9781316662205.

[195] S. Gupta and A. Saxena, Eds., *The Role of Topology in Materials*, vol. 189. Cham: Springer International Publishing, 2018. doi: 10.1007/978-3-319-76596-9.

[196] X.-L. Qi, R. Li, J. Zang, and S.-C. Zhang, "Inducing a Magnetic Monopole with Topological Surface States," *Science*, vol. 323, no. 5918, Art. no. 5918, Feb. 2009.

[197] X.-L. Qi, T. L. Hughes, and S.-C. Zhang, "Topological field theory of time-reversal invariant insulators," *Phys. Rev. B*, vol. 78, no. 19, p. 195424, Nov. 2008, doi: 10.1103/PhysRevB.78.195424.

[198] B.-S. Lu, "The Casimir Effect in Topological Matter," *Universe*, vol. 7, p. 237, Jul. 2021.

[199] J. A. Crosse, S. Fuchs, and S. Y. Buhmann, "Electromagnetic Green's function for layered topological insulators," *Phys. Rev. A*, vol. 92, no. 6, p. 063831, Dec. 2015, doi: 10.1103/PhysRevA.92.063831.

[200] S. Fuchs, J. A. Crosse, and S. Y. Buhmann, "Casimir-Polder shift and decay rate in the presence of nonreciprocal media," *Phys. Rev. A*, vol. 95, no. 2, p. 023805, Feb. 2017, doi: 10.1103/PhysRevA.95.023805.

[201] B.-S. Lu, K. Z. Arifa, and X. R. Hong, "Spontaneous emission of a quantum emitter near a Chern insulator: Interplay of time-reversal symmetry breaking and Van Hove singularity," *Phys. Rev. B*, vol. 101, no. 20, Art. no. 20, May 2020, doi: 10.1103/PhysRevB.101.205410.

[202] T. Cysne, W. J. M. Kort-Kamp, D. Oliver, F. A. Pinheiro, F. S. S. Rosa, and C. Farina, "Tuning the Casimir-Polder interaction via magneto-optical effects in graphene," *Phys. Rev. A*, vol. 90, no. 5, p. 052511, Nov. 2014, doi: 10.1103/PhysRevA.90.052511.

[203] W. Fang, G.-X. Li, J. Xu, and Y. Yang, "Enhancement of long-distance Casimir-Polder interaction between an excited atom and a cavity made of metamaterials," *Opt. Express*, vol. 27, no. 26, Art. no. 26, Dec. 2019, doi: 10.1364/OE.27.037753.

[204] V. N. Marachevsky and Y. M. Pis'mak, "Casimir-Polder effect for a plane with Chern-Simons interaction," *Phys. Rev. D*, vol. 81, no. 6, p. 065005, Mar. 2010, doi: 10.1103/PhysRevD.81.065005.

[205] G. S. Agarwal, "Anisotropic Vacuum-Induced Interference in Decay Channels," *Phys. Rev. Lett.*, vol. 84, no. 24, pp. 5500–5503, Jun. 2000, doi: 10.1103/PhysRevLett.84.5500.

[206] E. Lassalle *et al.*, "Long-lifetime coherence in a quantum emitter induced by a metasurface," *Phys. Rev. A*, vol. 101, no. 1, p. 013837, Jan. 2020, doi: 10.1103/PhysRevA.101.013837.

[207] G. Gómez-Santos, "Thermal van der Waals interaction between graphene layers," *Phys. Rev. B*, vol. 80, p. 245424, Dec. 2009.

[208] G. Bimonte, G. L. Klimchitskaya, and V. M. Mostepanenko, "How to observe the giant thermal effect in the Casimir force for graphene systems," *Phys. Rev. A*, vol. 96, p. 012517, Jul. 2017.

[209] G. L. Klimchitskaya and V. M. Mostepanenko, "Impact of graphene coating on the atom-plate interaction," *Phys. Rev. A*, vol. 89, p. 062508, Jun. 2014.

[210] M. Kardar, T. Emig, and S. J. Rahi, "Constraints on Stable Equilibria with Fluctuation-Induced (Casimir) Forces," *Phys. Rev. Lett.*, vol. 105, no. 7, p. 070404, Aug. 2010, doi: 10.1103/PhysRevLett.105.070404.

[211] S. Ribeiro, S. Yoshi Buhmann, T. Stielow, and S. Scheel, "Casimir-Polder interaction from exact diagonalization and surface-induced state mixing," *EPL Europhys. Lett.*, vol. 110, no. 5, p. 51003, Jun. 2015, doi: 10.1209/0295-5075/110/51003.

[212] H. Kübler, J. P. Shaffer, T. Baluktsian, R. Löw, and T. Pfau, "Coherent excitation of Rydberg atoms in micrometre-sized atomic vapour cells," *Nat. Photonics*, vol. 4, no. 2, Art. no. 2, Feb. 2010, doi: 10.1038/nphoton.2009.260.

[213] H. Failache, L. Amy, S. Villalba, L. Lenci, A. Laliotis, and A. Lezama, "Spectroscopy of Atoms Confined to Micrometric Pores in Glass," 2016, p. LTu2B.5. doi: 10.1364/LAOP.2016.LTu2B.5.

[214] S. Villalba, H. Failache, A. Laliotis, L. Lenci, S. Barreiro, and A. Lezama, "Rb optical resonance inside a random porous medium," *Opt. Lett.*, vol. 38, no. 2, pp. 193–195, Jan. 2013, doi: 10.1364/OL.38.000193.

[215] J. Sheng, Y. Chao, and J. P. Shaffer, "Strong Coupling of Rydberg Atoms and Surface Phonon Polaritons on Piezoelectric Superlattices," *Phys. Rev. Lett.*, vol. 117, no. 10, Aug. 2016, doi: 10.1103/PhysRevLett.117.103201.

[216] F. Ripka, H. Kübler, R. Löw, and T. Pfau, "A room-temperature single-photon source based on strongly interacting Rydberg atoms," *Science*, vol. 362, no. 6413, pp. 446–449, Oct. 2018, doi: 10.1126/science.aau1949.

[217] G. Epple *et al.*, "Rydberg atoms in hollow-core photonic crystal fibres," *Nat. Commun.*, vol. 5, Jun. 2014, doi: 10.1038/ncomms5132.

[218] K. S. Rajasree, T. Ray, K. Karlsson, J. L. Everett, and S. N. Chormaic, "Generation of cold Rydberg atoms at submicron distances from an optical nanofiber," *Phys. Rev. Res.*, vol. 2, no. 1, p. 012038, Feb. 2020, doi: 10.1103/PhysRevResearch.2.012038.

[219] X. Alauze, A. Bonnin, C. Solaro, and F. P. D. Santos, "A trapped ultracold atom force sensor with a $\upmu$m-scale spatial resolution," *New J. Phys.*, vol. 20, no. 8, p. 083014, Aug. 2018, doi: 10.1088/1367-2630/aad716.

[220] C. M. Soukoulis and M. Wegener, "Past achievements and future challenges in the development of three-dimensional photonic metamaterials," *Nat. Photonics*, vol. 5, no. 9, pp. 523–530, Sep. 2011, doi: 10.1038/nphoton.2011.154.

[221] G. Bimonte, T. Emig, R. L. Jaffe, and M. Kardar, "Spectroscopic probe of the van der Waals interaction between polar molecules and a curved surface," *Phys. Rev. A*, vol. 94, no. 2, Aug. 2016, doi: 10.1103/PhysRevA.94.022509.

[222] P. Thiyam *et al.*, "Anisotropic contribution to the van der Waals and the Casimir-Polder energies for $CO_2$ and $CH_4$ molecules near surfaces and thin films," *Phys. Rev. A*, vol. 92, no. 5, Nov. 2015, doi: 10.1103/PhysRevA.92.052704.

[223] M. Antezza, I. Fialkovsky, and N. Khusnutdinov, "Casimir-Polder force and torque for anisotropic molecules close to conducting planes and their effect on $CO_2$," *Phys. Rev. B*, vol. 102, no. 19, p. 195422, Nov. 2020, doi: 10.1103/PhysRevB.102.195422.

[224] J. Fiedler *et al.*, "Impact of effective polarisability models on the near-field interaction of dissolved greenhouse gases at ice and air interfaces," *Phys. Chem. Chem. Phys.*, vol. 21, no. 38, pp. 21296–21304, 2019, doi: 10.1039/C9CP03165K.

[225] A. Shih, D. Raskin, and P. Kusch, "Investigation of the interaction potential between a neutral molecule and a conducting surface," *Phys. Rev. A*, vol. 9, no. 2, pp. 652–662, Feb. 1974, doi: 10.1103/PhysRevA.9.652.

[226] C. Wagner *et al.*, "Non-additivity of molecule-surface van der Waals potentials from force measurements," *Nat. Commun.*, vol. 5, p. 5568, Nov. 2014, doi: 10.1038/ncomms6568.

[227] C. Brand *et al.*, "A Green's function approach to modeling molecular diffraction in the limit of ultra-thin gratings: A Green's function approach to modeling molecular diffraction," *Ann. Phys.*, vol. 527, no. 9–10, pp. 580–591, Oct. 2015, doi: 10.1002/andp.201500214.

[228] J. Lukusa Mudiayi *et al.*, "Linear Probing of Molecules at Micrometric Distances from a Surface with Sub-Doppler Frequency Resolution," *Phys. Rev. Lett.*, vol. 127, no. 4, p. 043201, Jul. 2021, doi: 10.1103/PhysRevLett.127.043201.